\documentclass{elsart}
\usepackage{graphicx,amssymb}
\usepackage{algorithmic}
\usepackage{algorithm}
\usepackage[switch,pagewise]{lineno}
\usepackage[usenames]{color}
\usepackage{longtable}
\usepackage{booktabs}
\usepackage{url}
\usepackage{subfigure}
\usepackage{threeparttable}
\usepackage{mathrsfs}
\usepackage{amsfonts}

\begin{document}
\begin{frontmatter}
\title{A Three-Phase Search Approach for the Quadratic Minimum Spanning Tree Problem}
\author{Zhang-Hua Fu}
\ead{fu@info.univ-angers.fr} \and
\author{Jin-Kao Hao\corauthref{cor}}
\corauth[cor]{Corresponding author.} \ead{hao@info.univ-angers.fr}
\address{LERIA, Universit\'{e} d'Angers \\ 2 Boulevard Lavoisier, 49045 Angers Cedex 01, France}

\maketitle

\begin{abstract}

Given an undirected graph with costs associated with each edge as well as each pair of edges, the \textsl{quadratic minimum spanning tree problem (QMSTP)} consists of determining a spanning tree of minimum total cost. This problem can be used to model many real-life network design applications, in which both routing and interference costs should be considered. For this problem, we propose a three-phase search approach named TPS, which integrates 1) a descent-based neighborhood search phase using two different move operators to reach a local optimum from a given starting solution, 2) a local optima exploring phase to discover nearby local optima within a given regional search area, and 3) a perturbation-based diversification phase to jump out of the current regional search area. Additionally, we introduce dedicated techniques to reduce the neighborhood to explore and streamline the neighborhood evaluations. Computational experiments based on hundreds of representative benchmarks show that TPS produces highly competitive results with respect to the best performing approaches in the literature by improving the best known results for 31 instances and matching the best known results for the remaining instances only except two cases. Critical elements of the proposed algorithms are analyzed.

\end{abstract}
\end{frontmatter}

\renewcommand{\baselinestretch}{1.0}\small\normalsize
\section{Introduction}

Let $G=(V,E)$ be a connected undirected graph with $|V|=n$ vertices and $|E|=m$ edges. Let $c: E \rightarrow \mathbb{R}$ be a linear cost function for the set of edges and $q: E\times E \rightarrow \mathbb{R}$ be a quadratic cost function to weight each pair of edges (without loss of generality, assume $q_{ee}=0$ for all $e\in E$). The \textsl{quadratic minimum spanning tree problem (QMSTP)} requires to determine a spanning tree $T=(V,X)$, so as to minimize its total cost $F(T)$, i.e., the sum of the linear costs plus the quadratic costs. Naturally, this problem can be formulated as follows (Cordone \& Passeri, 2012):

\begin{equation} \label{EqObjective}
\textrm{Minimize} \ F(T)=\sum_{e \in E}c_ex_e+\sum_{e\in E}\sum_{f\in E}q_{ef}x_ex_f,
\end{equation}

\begin{equation} \label{EqTree}
\textrm{subject to} \\ \sum_{e \in E}x_e=n-1,
\end{equation}

\begin{equation} \label{EqConnect}
 \sum_{e \in E(S)}x_e\le |S|-1, S\subset V: |S| \ge 3,
\end{equation}

\begin{equation}
 x_{e}\in \{0,1\}, e\in E.
\end{equation}

where $x_e=1$ if edge $e$ belongs to the solution, $x_e=0$ otherwise. $S$ is a subset of $V$ and $E(S)$ denotes the set of edges with both end vertices in $S$.

As an extension of the classical \textsl{minimum spanning tree problem (MST)} in graphs, the QMSTP has various practical applications in network design problems, where the linear function models the cost to build or use edges, while the quadratic function models interference costs between pairs of edges. For example, in transportation, telecommunication or oil supply networks, the linear function represents the costs for building each road, communication link or pipe, and the quadratic function represents the extra costs needed for transferring from one road (link, pipe) to another one. Normally, the interference costs are limited to pairs of adjacent edges (Maia, Goldbarg, \& Goldbarg, 2013; Pereira, Gendreau, \& Cunha, 2013), but in some special cases, the interference costs also exist between any pair of edges, especially for situations where the topology has little relation to the physical layout. As discussed in (Assad \& Xu, 1992; \"{O}ncan \& Punnen, 2010; Palubeckis, Rubliauskas, \& Targamadz\`{e}, 2010), the QMSTP has several equivalent formulations such as the \textsl{stochastic minimum spanning tree problem (SMSTP)}, the \textsl{quadratic assignment problem (QAP)}, and the \textsl{unconstrained binary quadratic optimization problem (UBQP)}.

During the last two decades, the QMSTP has been extensively investigated and many heuristic and exact approaches have been proposed. Since the QMSTP is $\mathcal{NP}$-hard and is even difficult to approximate (Xu, 1995), exact methods are often applied only to solve very small instances. For larger instances, heuristics are preferred to obtain feasible solutions within a reasonable time.

As for exact methods, Assad and Xu (Assad \& Xu, 1992; Xu, 1995) propose a Lagrangian  branch-and-bound method. \"{O}ncan and Punnen (2010) combine the Lagrangian relaxation scheme with an extended formulation of valid inequalities to obtain tighter bounds. Cordone and Passeri (2012) re-implement the Lagrangian branch-and-bound procedure in (Assad \& Xu, 1992) with some improvements. Very recently, several exact algorithms are proposed for solving other closely related variants. Respectively, Buchheim and Klein (2013a; 2013b) consider a special case of the QMSTP with exactly one quadratic term in the objective function, and propose a branch and bound approach to solve it. The description of this problem is further completed in (Fischer \& Fischer, 2013). Pereira, Gendreau, and Cunha (2013) propose a dynamic column and row generation based linear programming approach, to obtain strong lower bounds for the QMSTP with adjacency costs, for which the interference costs are only limited to adjacent edges.

On the other hand, heuristics which attempt to obtain near-optimal results within a reasonable time become the main approaches for solving the \textit{QMSTP}, especially for large instances. For instance, two greedy algorithms are proposed in (Xu, 1984; Assad \& Xu, 1992; Xu, 1995). Several genetic algorithms are implemented by Zhou and Gen (1998) and tested on instances with up to 50 vertices, showing that their algorithms dominate the above greedy algorithms. Another evolutionary algorithm is proposed for a fuzzy variant of the QMSTP (Gao, Lu, \& Li, 2004; Gao \& Lu, 2005), using the Pr\"{u}fer number to encode a spanning tree. Soak, Corne, and Ahn (2005; 2006) report remarkable results with an evolutionary algorithm using an edge-window-decoder strategy. In addition to these early methods, even more heuristics have been proposed in recent years, mostly based on local search. For example, the Tabu Thresholding algorithm (\"{O}ncan \& Punnen, 2010) alternatively performs local search and random moves. In (Palubeckis, Rubliauskas, \& Targamadz\`{e}, 2010), an iterated tabu search (ITS) is proposed and compared to a multi-start simulated annealing algorithm and a hybrid genetic algorithm, showing that ITS performs the best. An artificial bee colony algorithm is developed by Sundar and Singh (2010). Cordone and Passeri (2012) adopt a novel data structure and updating technique to reduce the amortized time at each iteration of neighborhood exploration from $O(mn^2)$ to $O(mn)$, based on which they further propose a tabu search (TS) algorithm and report a number of improved results over previous best known results. Very recently, Lozano et al. (2013) propose an iterated greedy (IG) and a strategic oscillation (SO) heuristic, and combine them with the ITS (Palubeckis, Rubliauskas, \& Targamadz\`{e}, 2010) algorithm to obtain a powerful hybrid algorithm named HSII. In addition to the standard QMSTP, for the variant only with adjacency costs, Maia, Goldbarg, \& Goldbarg (2013) propose a Pareto local search algorithm and adapt the 108 instances in (Cordone \& Passeri, 2012) as benchmarks to evaluate the proposed algorithm.

In this paper, we are interested in the QMSTP and propose a three-phase search approach named TPS, which uses a randomized constructive procedure for solution initialization, and integrates a descent-based neighborhood search phase based on two different move operators to reach a local optimum from any given starting solution, a local optima exploring phase to discover different local optima within a given regional search area, and a perturbation-based diversification phase to jump out of the current search area and move to unexplored search areas. As a supplementary technique, a fast examination technique is implemented to identify and discard hopeless neighboring solutions, so as to avoid useless computations. Computational results based on 7 sets of 659 representative benchmarks demonstrate that TPS yields highly competitive results with respect to the best performing heuristics, showing its effectiveness and efficiency for solving the QMSTP. In particular, for the 630 conventional QMSTP instances mostly with unknown optimal results, TPS improves the best known results for 31 instances and matches the best known results for the remaining instances only except two cases. For the set of the 29 instances transformed from the \textsl{Quadratic Assignment Problem} which are known to be difficult for existing QMSTP algorithms, TPS consistently solve them to optimality within a very short time.

The rest of this paper is organized as follows: Section \ref{SecApproach} describes the proposed approach. After providing the computational results in Section \ref{SecResults}, Section \ref{SecAnalyze} analyzes the influence of several important ingredients of the proposed algorithm. Finally, conclusions from this study are drawn in Section \ref{SecConclude}.

\section{A three-phase search approach for the QMSTP}
\label{SecApproach}

\subsection{General framework}
\label{SubsecFramework}

The proposed three-phase search approach TPS for the QMSTP is outlined in Algorithm \ref{AlgorithmOutline}, which is composed of several subroutines. Respectively, \textsl{Init\_Solution} is used to randomly generate an initial starting solution. The first search phase is ensured by \textsl{Descent\_Neighborhood\_Search} which is a descent-based neighborhood search procedure and aims to reach a local optimal solution from a given starting solution. The second search phase \textsl{Explore\_Local\_Optima} is used to discover nearby local optima of better quality within the current regional search space. If no further improvement can be attained, the search turns into a diversified perturbation phase \textsl{Diversified\_Perturb}, which strongly perturbs the incumbent solution to jump out of the current regional search area in order to displace the search into a new area. After that, the search enters into a new round of \textsl{Descent\_Neighborhood\_Search} and \textsl{Explore\_Local\_Optima} search phases again. This process is iterated until the terminal criterion is met.

\begin{figure}[!hbp]
\includegraphics[width=12cm]{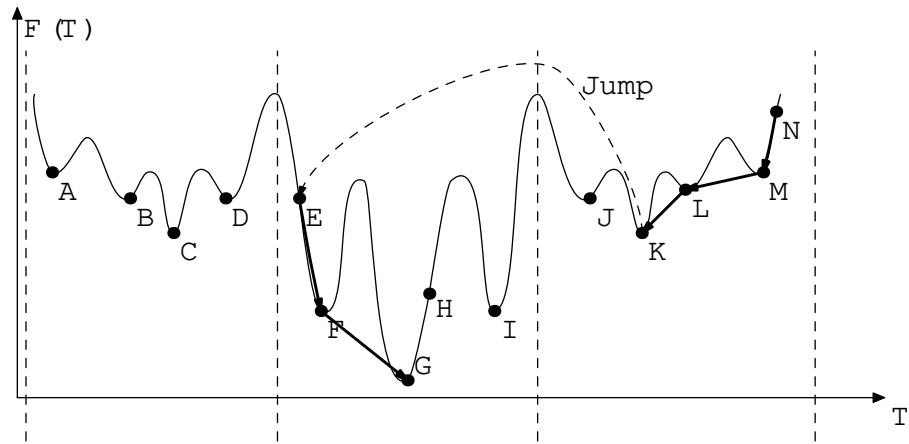}
\centering\caption{Procedure of searching a high-quality feasible solution of the QMSTP}
\label{FigSearchProcedure}
\end{figure}

\begin{algorithm}[!hbp]
\footnotesize
 \caption{Framework of the Proposed Approach for the QMSTP} \label{AlgorithmOutline}
 \begin{algorithmic}[h]
   \REQUIRE Graph $G(V,E)$, linear function $E \rightarrow \mathbb{R}$, quadratic function $E\times E \rightarrow \mathbb{R}$
   \ENSURE The best solution found
   \STATE /* Construct an initial solution, Section \ref{SubsecInit} */
   \STATE $T \leftarrow Init\_Solution()$
   \STATE /* Find a local optimum with the descent-based neighborhood search phase, Section \ref{SubsecNS} */
   \STATE $T \leftarrow Descent\_Neighborhood\_Search(T)$
   \STATE /* Explore nearby local optima of better quality, Alg. \ref{AlgorithmIntensiveSearch} and Section \ref{SubsecIntensiveSearch} */
   \STATE $T \leftarrow Explore\_Local\_Optima(T)$
   \STATE /* $T^{best}$ records the best solution found so far */
   \STATE $T^{best} \leftarrow T$
   \WHILE{The terminal criterion is not met}
      \STATE /* Strongly perturb the incumbent solution to a new solution, Section \ref{SubsecDiverSearch} */
      \STATE $T \leftarrow Diversified\_Perturb(T)$
      \STATE /* Call the neighborhood search phase to reach a local optimum again */
      \STATE $T \leftarrow Descent\_Neighborhood\_Search(T)$
      \STATE /* Enter the local optima exploring phase again */
      \STATE $T \leftarrow Explore\_Local\_Optima(T)$
      \STATE /* Update $T^{best}$ if an improved solution is found */
      \IF{$F(T)<F(T^{best})$}
        \STATE $T^{best} \leftarrow T$
      \ENDIF
   \ENDWHILE
   \RETURN $T^{best}$
\end{algorithmic}
\end{algorithm}

Fig. \ref{FigSearchProcedure} illustrates the idea followed by the TPS procedure, where $X$-axis indicates all the feasible solutions $T$, and $Y$-axis indicates the corresponding objective values $F(T)$. As shown in Fig. \ref{FigSearchProcedure}, $A,B,C,D,F,G,I,J,K,L,M$ are local optima of different qualities, while $E,H,N$ being feasible solutions. Starting from a randomly generated initial solution, say $N$, the search calls \textsl{Descent\_Neighborhood\_Search} to reach a first local optimum $M$, and then uses the \textsl{Explore\_Local\_Optima} search phase to discover nearby local optima $L$ and $K$. At this point, the \textsl{Diversified\_Perturb} phase is executed to jump from $K$ to a faraway enough solution $E$, which is subsequently optimized by \textsl{Descent\_Neighborhood\_Search} ($E\rightarrow F$) and \textsl{Explore\_Local\_Optima} ($F\rightarrow G$), to obtain a high-quality solution $G$.

In its most general form, the proposed procedure can be considered as an iterated local search algorithm (Lourenco, Martin, St\"{u}tzle, 2003). Nevertheless, TPS is distinguished by its local optima exploring phase and its perturbation mechanisms which share similarities with breakout local search (Benlic \& Hao, 2013a; 2013b; 2013c; Fu \& Hao, 2014).

\subsection{Solution presentation}

Like the compact tree representation used in (Cordone \& Passeri, 2012; Fu \& Hao, 2014), we uniquely represent each feasible solution $T$ as a rooted tree (with vertex 1 fixed as the root vertex, being different from (Cordone \& Passeri, 2012) where the root changes dynamically during the search process), corresponding to a one-dimensional vector $T=\{t_i, i\in V \}$, where $t_i$ denotes the parent vertex of vertex $i$ only except the root vertex 1 (let $t_1=null$). Inversely, given a vector $T=\{t_i, i\in V \}$, the corresponding solution tree can be easily reconstructed.

\subsection{Initialization}
\label{SubsecInit}

TPS requires an initial solution to start its search. Moreover, given its stochastic nature, multiple runs of TPS from different initial solutions are typically applied to find the best possible solutions for a problem instance. For the QMSTP, starting from an empty solution $T$ containing only the root vertex and no edge, we iteratively select at random one edge from $E$ and add it to $T$ (without leading to any closed loop), until $n-1$ such edges are added, meaning that a feasible initial solution is generated. In the rest of the paper, a feasible tree will be noted by $T=(V,X)$ where $V$ and $X \subseteq E$ are respectively the vertex set of the graph and the set of edges of the tree.

\subsection{Descent-based neighborhood search phase}
\label{SubsecNS}

As the basis of the proposed approach, a descent-based neighborhood search phase \textsl{Descent\_Neighborhood\_Search} is used to reach a local optimum from a given starting solution $T=(V,X)$. For this, we develop two different move operators to generate neighboring solutions.

\begin{figure}[!hbp]
\centering{\includegraphics{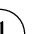}}
\centering{\includegraphics{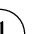}}\\
\centering{\includegraphics{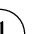}}
\centering{\includegraphics{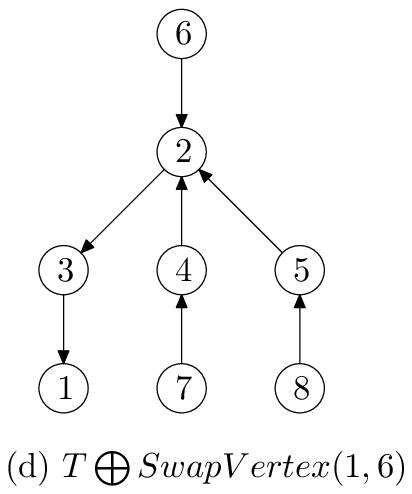}}
\centering\caption{Move operators for generating neighboring solutions}
\label{FigMoveOperator}
\end{figure}

\begin{enumerate}
  \item The first one is the conventional \textsl{swap-edge move operator} inherited from (Palubeckis, Rubliauskas, \& Targamadz\`{e}, 2010; Cordone \& Passeri, 2012; Lozano et al., 2013). This operator first adds to $X$ one of the $m-n+1$ unused edges $e\in E\backslash X$, thus closing a loop $L_e$ of $|L_e|\le n$ edges, and then removes an edge $f$ from $L_e\backslash{e}$, to obtain a feasible neighboring solution denoted by $T \bigoplus SwapEdge(e,f)$. The corresponding difference of the objective function (also called move gain) is denoted by $\delta_{ef}$.
  \item The above move operator swaps only one pair of edges. It is tempting to introduce a move operator by swapping two pairs of edges to obtain an enlarged neighborhood. Nevertheless, such a move operator induces a neighborhood with a total of $O(m^2n^2)$ neighboring solutions, which is extremely expensive for neighborhood examination. To control the size of the neighborhood, we develop a restricted \textsl{swap-vertex move operator} as follows. Let $V^1 \subseteq V$ denote the subset containing all the vertices with degree equal to 1 (including all the leaf vertices and possibly the root vertex), and for each vertex $i \in V^1$, let $r_i$ denote the related vertex, i.e., the vertex connected to $i$. Then, for each pair of vertices $i, j \in V^1$ with $r_i \ne r_j$ and $\{i,r_j\} \in E, \{j,r_i\} \in E$, a feasible neighboring solution denoted by $T \bigoplus SwapVertex(i,j)$ could be generated by swapping vertices $i$ and $j$, leading to a difference $\delta_{ij}$ of the objective function. Note that, if we denote edges $\{i,r_j\}$, $\{j,r_i\}$, $\{i,r_i\}$, $\{j,r_j\}$ by $e1$, $e2$, $f1$, $f2$ respectively, $SwapVertex(i,j)$ is indeed equivalent to $SwapEdge(e1,f1) \bigoplus SwapEdge(e2,f2)$. Clearly,  $SwapVertex(i,j)$ defines a neighborhood whose size is bounded by  $O(n^2)$.
\end{enumerate}

For example, Fig. \ref{FigMoveOperator} illustrates several neighboring solutions generated by the above move operators. Respectively, based on the original solution $(a)$, solution $(b)$ is generated by adding edge $\{5,7\}$ and subsequently deleting edge $\{2,4\}$, while solutions $(c)$ and $(d)$ are obtained by swapping the related vertices corresponding to vertices $6$ and $8$, $1$ and $6$ respectively. Note that, after applying the move operators, the parent-child relationship of some vertices should be reversed (for example, the relationship between vertices 2 and 3 in sub-figure (d)), to ensure that vertex 1 is always fixed as the root.

Based on these move operators ($SwapEdge(e,f)$ and $SwapVertex(i,j)$), two different neighborhoods $N1$ and $N2$ are defined as follows:

\begin{equation} \label{EqMove}
\begin{array}{ll}
N1=\{T \bigoplus SwapEdge(e,f)\ |\ e\in E\backslash X,\ f\in L_e\backslash{e} \},\\
N2=\{T \bigoplus SwapVertex(i,j) \ |\ i, j\in V^1, r_i \ne r_j, \{i,r_j\} \in E, \{j,r_i\} \in E \}.\\
\end{array}
\end{equation}

where  $T=(V,X)$ is a feasible solution, and $T \bigoplus OP$ designates the neighboring solution obtained by applying $SwapEdge$ or $SwapVertex $ to $T$.

Typically, \textsl{Descent\_Neighborhood\_Search} examines (in random order) the feasible neighboring solutions belonging to $N1 \bigcup N2$ (i.e. the union of the two neighborhoods) and applies the first met improving one (with $\delta_{ef}<0$ or $\delta_{ij}<0$) to replace the incumbent solution. This process is iterated until no such improving solution exists in the neighborhood, meaning that a local optimum is reached.

Additionally, one observes that $|N1|=O(m-n-1)\times O(|L_e|)\le O(mn)\le O(n^3)$, and $|N2|=O(|V^1|^2)\le O(n^2)$, being statistically much less than $|N1|$. To ensure a fast exploration of neighborhood $N1$, in the following subsection, we develop a pre-estimation technique which is able to identify and discard a large number of useless $SwapEdge(e,f)$ moves, consequently to reduce the computational complexity needed for \textsl{Descent\_Neighborhood\_Search}.

\subsection{Fast examination technique}
\label{SubsecFastEvalute}

Like in (Cordone \& Passeri, 2012), we maintain a vector $D$, whose values indicate the actual or potential contribution of each edge $g\in E$ to the overall cost of the incumbent solution $T=(V,X)$.

\begin{equation} \label{EqVctor}
 D_g=c_g+\sum_{h\in X}(q_{gh}+q_{hg}), \forall g\in E.
\end{equation}

With this vector, for each of the $O(mn)$ possible swap-edge moves $SwapEdge(e,f)$ corresponding to the neighborhood $N1$, the objective function varies by (Cordone \& Passeri, 2012):

\begin{equation} \label{EqDeltaEF}
 \delta_{ef}=D_e-D_f-q_{ef}-q_{fe}.
\end{equation}

which can be calculated in constant time $O(1)$. Additionally, after performing the chosen move $SwapEdge(e,f)$, vector $D$ is updated as follows (Cordone \& Passeri, 2012):

\begin{equation} \label{EqUpdateEF}
 D_g\leftarrow D_g+q_{ge}+q_{eg}-q_{gf}-q_{fg}, \ \forall g\in E.
\end{equation}

which takes a total time of $O(m)$. Clearly, the overall complexity for exploring the first neighborhood $N1$ at each iteration is $O(mn)\times O(1)+O(m)=O(mn)$.

Similarly, since each of the $O(n^2)$ possible swap-vertex moves $SwapVertex(i,j)$ (i.e. neighborhood $N2$) is equivalent to $SwapEdge(e1,f1) \bigoplus SwapEdge(e2,f2)$, where $e1$, $e2$, $f1$, $f2$ denote edges $\{i,r_j\}$, $\{j,r_i\}$, $\{i,r_i\}$, $\{j,r_j\}$ respectively, the difference of the objective function is obtained by:

\begin{equation} \label{EqDeltaIJ}
\begin{array}{ll}
\delta_{ij}=D_{e1}+D_{e2}-D_{f1}-D_{f2}+q_{e1e2}+q_{e2e1}+q_{f1f2}+q_{f2f1}\\-q_{e1f1}-q_{f1e1}-q_{e1f2}-q_{f2e1}-q_{e2f1}-q_{f1e2}-q_{e2f2}-q_{f2e2}.
\end{array}
\end{equation}

where each term can be evaluated in constant time $O(1)$. Then, vector $D$ is updated in $O(m)$ as follows:

\begin{equation} \label{EqUpdateIJ}
D_g \leftarrow D_g+q_{ge1}+q_{e1g}+q_{ge2}+q_{e2g}-q_{gf1}-q_{f1g}-q_{gf2}-q_{f2g}, \ \forall g\in E.
\end{equation}

Clearly, the computational complexity needed for exploring $N2$ at each iteration is at most $O(n^2)\times O(1)+O(m)=O(n^2)$.

Furthermore, we attempt to reduce the computational time needed for the examination of neighborhood $N1$, which is the most expensive part of the first search phase. As mentioned above, at each iteration of \textsl{Descent\_Neighborhood\_Search}, up to $O(mn)$ legal swap-edge moves are possible. However, many of these moves are definitely hopeless since no improvement over the incumbent solution can be gained. Since \textsl{Descent\_Neighborhood\_Search} only accepts improved solutions with $\delta_{ef}<0$, it is interesting to identify these hopeless moves with $\delta_{ef}\ge 0$ and discard them directly to avoid irrelevant computations.

Based on this idea, we develop a pre-estimation criterion as follows. Let $\gamma=Max\{D_g, g\in X\}$ denote the maximum cost value of $D_g$ of all the edges $g$ belonging to the incumbent solution $T=(V,X)$, and let $\lambda=Max\{q_{hk}+q_{kh},\ h,\ k \in E\}$ denote the maximum possible value of quadratic costs between any pair of edges. Note that $\gamma$ is a variable which should be updated at each iteration, within an amount of $O(|X|)=O(n)$ extra time, while $\lambda$ is a constant. Then, it is clear that, for each edge $e\in E\backslash X$, if we add it to $X$, the objective function would increase by $D_e$. At this point, one can observe that no matter which edge $f\in L_e \backslash e$ we choose to remove from $X$, the decreased cost is strictly bounded within $\gamma+\lambda$. Obviously, if $D_e-\gamma-\lambda \ge 0$, it means that all the possible moves $SwapEdge(e,f), f\in L_e \backslash e$ lead to a solution no better than the incumbent solution $T$. In other words, it is definitely impossible to obtain an improved solution by exchanging $e$ against any other edge belonging to the incumbent solution. Consequently, we can directly discard all these moves to avoid useless evaluations, thus reducing the computation time.

While exploring the neighboring solutions belonging to neighborhood $N1$, for each edge $e\in E \backslash X$, we first use the above pre-estimation criterion to check if it is possible to gain any improvement by exchanging $e$ against some other edge $f\in L_e \backslash e$. If this is not the case, we discard all the moves involving $e$ and skip to the next edge in $E \backslash X$. Otherwise, we evaluate one by one the possible legal moves $SwapEdge(e,f), f\in L_e \backslash e$ to identify an improving neighboring solution. As shown in Section \ref{SecAnalyze}, this fast examination technique allows the algorithm to identify and discard a high number of hopeless moves, accelerating considerably the neighborhood exploration without any sacrifice of solution quality.

\subsection{Local optima exploring phase}

\label{SubsecIntensiveSearch}

Obviously, the \textsl{Descent\_Neighborhood\_Search} procedure described in Section \ref{SubsecNS} alone cannot go beyond the achieved local optimum. In order to be able to discover nearby local optima which are possibly of better quality and to intensify the search in a given regional search space, we develop a local optima exploring (\textsl{Explore\_Local\_Optima}) phase (Algorithm \ref{AlgorithmIntensiveSearch}). The local optima exploring phase is based on two directed perturbation operators (denoted by \textit{Directed\_Perturb}). Inspired by the idea of breakout local search (Benlic \& Hao, 2013a; 2013b; 2013c), these directed perturbation operators rely on the tabu search principle (Glover \& Laguna, 1997), which favors the moves with the weakest deterioration of the objective function. Precisely, \textsl{Directed\_Perturb} takes one of the following two forms.

\begin{algorithm}[!hbp]
\footnotesize
 \caption{Local Optima Exploring (\textsl{Explore\_Local\_Optima}) Phase} \label{AlgorithmIntensiveSearch}
 \begin{algorithmic}[1]
   \REQUIRE The incumbent local optimal solution $T$
   \ENSURE The best found local optimal solution near $T$
   \STATE /* $T^{\sharp}$ records the best local optimum found during the current local optima exploring phase */
   \STATE $T^{\sharp} \leftarrow T$
   \STATE /* $\omega$ counts the number of consecutive non-improving rounds of directed perturbation followed by neighborhood search */
   \STATE $\omega\leftarrow 0$
   \STATE /* \textsl{Explore\_Local\_Optima} stops once  $\omega$ reaches its upper bound $\omega_{max}$ */
   \WHILE{$\omega<\omega_{max}$}
      \STATE /* Apply a directed perturbation operator to perturb $T$ */
      \STATE $T \leftarrow Directed\_Perturb(T)$
      \STATE /* Optimize the incumbent solution to a new local optimum */
      \STATE $T \leftarrow Descent\_Neighborhood\_Search(T)$
      \STATE /* If $T$ is better than $T^{\sharp}$, update $T^{\sharp}$ and reset $\omega$ ; Otherwise increase $\omega$ */
      \IF{$F(T)<F(T^{\sharp})$}
        \STATE $T^{\sharp} \leftarrow T$
        \STATE $\omega\leftarrow 0$
      \ELSE
        \STATE $\omega \leftarrow \omega+1$
      \ENDIF
   \ENDWHILE
   \RETURN $T^{\sharp}$
\end{algorithmic}
\end{algorithm}

\begin{enumerate}
  \item The \textsl{swap-edge directed perturbation} operator applies the swap-edge move operator (see Section \ref{SubsecNS}). For each edge $g\in E$, this perturbation operator saves in an array the last iteration $I_g$ when edge $g$ is added into or removed from the current solution. With this information, before exchanging edge $e \in E \backslash X$ and edge $f \in L_e\backslash e$, we check at first whether the current iteration index is larger than both $I_e+l_{in}$ and $I_f+l_{out}$, where $l_{in}$ and $l_{out}$ are parameters indicating the length of the prohibition, i.e., the tabu tenures (Glover \& Laguna, 1997). If this is not the case, the corresponding move \textsl{SwapEdge(e,f)} is marked tabu. This prohibition aims to avoid the inclusion of a recently removed edge or the removal of a recently included edge, unless the move meets the aspiration criterion, i.e., leading to a solution better than the overall best found solution. Otherwise, \textsl{SwapEdge(e,f)} is declared non-tabu. Typically, the perturbation operator examines all the non-tabu legal moves and iteratively applies the best legal move to the incumbent solution (no matter it leads to an improved solution or not), until a given number $L_{dir}$ (called perturbation strength) of such moves are performed.

  \item The \textsl{swap-vertex directed perturbation} operator is based on the swap-vertex move operator (see Section \ref{SubsecNS}). For each vertex $v\in V$, we save in an array the last iteration $I_v$ when vertex $v$ is swapped with some other vertex. Then, before swapping any pair of vertices $i,j\in V^1$, we check at first whether the current iteration index is larger than both $I_i+l_{swap}$ and $I_j+l_{swap}$, where $l_{swap}$ is a parameter indicating the tabu tenure. The moves satisfying this condition are marked non-tabu, while the others are declared tabu, unless they meet the same aspiration criterion used above. Similarly, this perturbation operator iteratively applies the best non-tabu move to the incumbent solution, until a given number $L_{dir}$ of such moves are performed.
\end{enumerate}

Given these two types of directed perturbation operators, the local optima exploring (\textsl{Explore\_Local\_Optima}) phase applies them probabilistically:  selecting the swap-edge perturbation with probability $p$ (parameter), and applying the swap-vertex directed perturbation with probability $1-p$. Then, the \textsl{Descent\_Neighborhood\_Search} phase is applied again to the perturbed incumbent solution to obtain a new local optimum. Typically, the local optima exploring phase alternates between \textsl{Directed\_Perturb} and \textsl{Descent\_Neighborhood\_Search}, until no further improvement is gained after $\omega_{max}$ consecutive such rounds ($\omega_{max}$ is fixed to be 5 in this paper), meaning that it is difficult to find better local optima within the current search region. At this point, the search turns into a diversified perturbation phase described below, in order to jump out of the current region.

\subsection{Diversified perturbation phase}
\label{SubsecDiverSearch}

The diversified perturbation phase aims to jump out of the current regional search area and displace the search to more distancing search areas, while retaining a certain degree of structure information of the incumbent solution. For this, we develop a diversified perturbation operator \textsl{Diversified\_Perturb}, which iteratively removes at random an edge $f$ from $T=(V,X)$ and subsequently adds the best feasible edge $e \in E\backslash X$ into $T$, without leading to any closed loop (to ensure the feasibility of the solution after insertion), until a given number $L_{div}$ (parameter for controlling the perturbation strength) of such perturbation moves are performed.

Given the definition of the \textsl{Directed\_Perturb} and \textit{Diversified\_Perturb} operators, we understand that they introduce different degrees of diversification to the search process. Indeed, with tabu principle, \textsl{Directed\_Perturb} modifies the incumbent solution more gradually and keeps the search within areas close to the incumbent solution. On the other hand, by random moves, \textsl{Diversified\_Perturb} may disrupt strongly the incumbent solution and leads the search to a completely new region. By combining these two different types of perturbations, it is expected that a better trade-off between intensification and diversification would be reached in the general search procedure.

Finally, in our TPS algorithm (see Algorithm \ref{AlgorithmOutline}), each time the \textsl{Diversified\_Perturb} phase is applied to modify the incumbent local optimum $T$, the resulting solution is subsequently optimized by \textsl{Descent\_Neighborhood\_Search} followed by \textsl{Explore\_Local\_Optima}. This process is repeated, until the terminal criterion is satisfied. Then, the best found solution $T^{best}$ is returned as the obtained solution.

\section{Experimental results}
\label{SecResults}

In order to evaluate the performance of our TPS algorithm\footnote{All the best solutions reported in this paper are available on \url{http://www.info.univ-angers.fr/pub/hao/qmstp.html}, the TPS source code will also be made available online upon the publication of the paper.}, which is coded in C, we test it on a large number of well-known benchmarks, and then compare the results with respect to previous state-of-the-art heuristics in the literature. For comparisons, we consider the objective function values, i.e., the consumed cost as our main evaluation criterion, and include the runtime for indicative purposes. For information, the TPS algorithm is executed on an Intel Xeon E5440 2.83 GHz processor and 2 GB RAM, while a 1.6 GHz Pentium IV processor is used in (Soak, Corne, \& Ahn, 2006), a 3 GHz Pentium IV CPU and 2 GB RAM in (\"{O}ncan \& Punnen, 2010), a 3.0 GHz core 2 duo system with 2 GB RAM in (Sundar \& Singh, 2010), a 3.0 GHz Intel core 2 duo in (Palubeckis, Rubliauskas, \& Targamadz\`{e}, 2010), a 2.6 GHz Intel Pentium Core 2 Duo E6700 and 2 GB RAM in (Cordone \& Passeri, 2012), a 3.2 GHz Intel processor and 12 GB RAM in (Lozano et al., 2013). One can observe that the clock frequency of our processor is about 80\% faster than the computer used in (Soak, Corne, \& Ahn, 2006), while being similar to the machines used in (\"{O}ncan \& Punnen, 2010; Sundar \& Singh, 2010; Palubeckis, Rubliauskas, \& Targamadz\`{e}, 2010; Cordone \& Passeri, 2012; Lozano et al., 2013).

\subsection{Benchmark instances}
\label{SubsecInstance}

Given the importance of the QMSTP, a large number of benchmark instances are generated for evaluation. To the best of our knowledge, the following benchmarks are currently available, which could be further classified into seven groups as follows \footnote{The \textit{CP} benchmarks can be downloaded from \url{ http://www.dti.unimi.it/cordone/research/qmst.html} and the RAND and SOAK instances are available at \url{http://sci2s.ugr.es/qmst/QMSTPInstances.rar}. The others can be provided on request to the authors (fu@info.univ-angers.fr or hao@info.univ-angers.fr).}.

\begin{itemize}
 \item Benchmark CP (Cordone \& Passeri, 2008) consists of 108 instances, with vertices number $n$ ranging from 10 to 50, and graph density $\rho=33\%$, $67\%$ or $100\%$. The linear costs and the quadratic costs are randomly distributed in [1,10] or [1,100].

  \item Benchmark OP1 (\"{O}ncan \& Punnen, 2010) consists of 480 complete graphs, with $n$=6-18, 20, 30, 50 respectively, each group having 30 instances. These instances are further divided into three subclasses:

\begin{enumerate}
	\item SYM: with linear costs uniformly distributed at random within [1,100], and quadratic ones within [1,20];
	\item VSYM: the linear costs are uniformly distributed at random in [1,10000], for the quadratic costs, each vertex is assigned with a value randomly distributed in [1,10] and the quadratic cost $q_{ef}$ is obtained by multiplying the four values associated with the end vertices of edges $e$ and $f$;
	\item ESYM: the vertices are randomly distributed in a square of side 100, then the linear costs are the Euclidean distances between the end vertices of each edge, and the quadratic costs are the Euclidean distances between the mid-points of the edges.
\end{enumerate}

  \item Benchmark SCA (Soak, Corne, \& Ahn, 2006) includes 6 complete graphs, with vertices number ranging from 50 to 100, by steps equal to 10. For each instance, the vertices are uniformly spread in a square of side 500, then the linear costs are the Euclidean distances between the vertices and the quadratic costs are uniformly distributed within [0,20].

  \item Benchmark SS (Sundar \& Singh, 2010) consists of 18 complete graphs with $n=$25, 50, 100, 150, 200 and 250 (each corresponds to 3 instances), the linear costs are uniformly distributed at random within [1,100] and the quadratic costs are randomly distributed within [1,20].

  \item Benchmark RAND (Lozano et al., 2013): This group consists of 9 large instances (with $n=$150, 200 or 250) recently generated by Lozano et al. (2013), with linear costs uniformly distributed in [1, 100], and quadratic ones uniformly distributed in [1, 20].

  \item Benchmark SOAK (Lozano et al., 2013): This group also includes 9 large instances (with $n=$150, 200 or 250), with vertices uniformly distributed at random on a 500$\times$500 grid. The edge costs are the integer Euclidean distances between any pair of vertices, and the quadratic ones are uniformly distributed between [1, 20].

  \item Benchmark QAP-QMSTP consists of 29 special QMSTP instances converted from the NUG (Nugent, Vollman, \& Ruml, 1968) and CHR (Christofides \& Benavent, 1989) benchmarks of the \textsl{Quadratic Assignment Problem (QAP)}, using a one-to-one transformation procedure between these two different problems (\"{O}ncan \& Punnen, 2010). Note that, although all the original QAP instances have already been solved to optimality by previous QAP algorithms (Burkard, Karisch, \& Rendl, 1997), they are difficult for existing QMSTP algorithms to reach the optimal solutions, due to the quite special problem structures after transformation. Even the best QMSTP algorithm misses 17 optimal solutions.
\end{itemize}

As mentioned above, the CP and OP1 groups contains a large number of small sized instances, with up to $50$ vertices. In fact, for all the 108 instances of group CP and almost all (476 out of 480) the instances of group OP1, our TPS algorithm can easily match the previous best known results. More importantly, for the left 4 instances of group OP1, TPS succeeds in finding several improved solutions. In addition to these two groups, the following five groups of instances are much more challenging, due to their large-scale problem sizes or special problem structures. To emphasize the effectiveness of TPS for solving challenging instances, in this paper, we just provide in detail the results obtained on the five groups of challenging instances, with respect to previous state-of-the-art heuristics, while summarizing the results on groups CP and OP1, to shorten the paper.

\subsection{Parameters}
\label{SubsecParameter}

As described in Section \ref{SecApproach}, TPS requires several parameters: the tabu tenures $l_{in}$, $l_{out}$, $l_{swap}$ used in the directed perturbation operators, the perturbation strength $L_{dir}$ and $L_{div}$, as well as the probability $p$ for selecting a directed perturbation operator. Generally, these parameters could be tuned with respect to each benchmark group given that the groups have different characteristics and structures. However, to show the efficiency and the robustness of the proposed approach, we uniformly adopt a fixed set of parameter values for all the test instances only except those of group QAP-QMSTP.

First, we observe in our preliminary experiments that the swap-edge based directed perturbation operator alone performs quite well for all the benchmarks except the group QAP-QMSTP. Thus we set $p=1$ (thus disable the swap-vertex directed perturbation operator, rending parameter $l_{swap}$ unnecessary). The tabu tenures $l_{in}$, $l_{out}$ are distributed at random within $[1,3]$, $[0.35n,0.45n]$, and the perturbation strengths $L_{dir}$, $L_{div}$ are distributed with random integers belonging to $[0.5n,n]$ and $[0.4n,0.6n]$ respectively.

Second, for the group QAP-QMSTP, we observe that the swap-vertex based directed perturbation operator is extremely useful (as further discussed in Section \ref{SubsecAnalyzeDirectedPerturb}), and the tabu mechanism plays a more important role for solving these instances. Following these observations, we set parameter $p=0.5$, and enlarge parameter $L_{dir}$ to $[5n,10n]$, while keeping the same values for the other parameters as above. Additionally, for the extra tabu tenure $l_{swap}$ used in the swap-vertex based directed perturbation operator, it is distributed at random within $[n,5n]$.

The stopping criterion of our algorithm is provided in the next subsection along with the presentation of the computational results of each benchmark group. Finally, one notices that there is no uniform terminal criterion used by the previous heuristics, various terminal conditions have been adopted by different researchers, for solving different instances.

\subsection{Results of the CP instances}
\label{SubsecResultsCP}

Among the 108 instances generated in (Cordone \& Passeri, 2008), tens of the largest ones have been selected to evaluate several algorithms, i.e., ITS (Palubeckis, Rubliauskas, \& Targamadz\`{e}, 2010), QMST-TS (Cordone \& Passeri, 2012), HSII (Lozano et al., 2013). Respectively, both ITS and QMST-TS solve each instance 10 times, each run continues until the previous best known solution is reached\footnote{The previous best known results for the 108 CP instances are available at \url{ http://www.dti.unimi.it/cordone/research/qmst.html}.}. Experimental results show that for all these selected instances, each run of ITS and QMST-TS can unexceptionally reach the best known solution, with a mean computing time (average of each independent run) ranging from less than 1 second to about 2 minutes. HSII also executed 10 independent times to solve each instance, with a cutoff time of 10 seconds for each run. However, for many test instances, HSII occasionally fails to match the previous best known results within the allowed time.

To evaluate the performance of our TPS algorithm on this set of 108 instances, we follow ITS and QMST-TS. In other words, we independently run TPS 10 times to solve each instance, each run continues until the best known result is reached. Our results show that, each TPS run unexceptionally succeeds in reaching the best known result, with an average time from less than 1 second to less than 2 minutes, indicating that TPS performs similarly with respect to ITS and QMST-TS for this group of small benchmarks. Since these instances are not challenging enough, we do not show our detailed results.

\subsection{Results of the \textit{OP1} instances}
\label{SubsecResultsOP1}

This group consists of three subclasses (SYM, ESYM, VSYM), each includes 160 instances, with $n$ ranging from 6 to 50 (a total of 480 instances). These benchmarks have been used to evaluate several previous algorithms, including two exact approaches, i.e., the refined Lagrangian lower bounding procedure in (\"{O}ncan \& Punnen, 2010) and the branch-and-bound algorithm QMST-BB in (Cordone \& Passeri, 2012), as well as two heuristics, i.e., the RLS-TT algorithm in (\"{O}ncan \& Punnen, 2010) and the tabu search algorithm QMST-TS in (Cordone \& Passeri, 2012). Nevertheless, the existing exact approaches can only solve small sized instances with up to $n=20$ vertices. For larger instances, only heuristics are able to produce feasible solutions within a reasonable time. In addition, RLS-TT just provides summarized results for these benchmarks, without giving detailed results for each instance. Unfortunately, some of the results reported by RLS-TT exhibit internal inconsistencies, probably due to typos mistakes, as pointed out in (Cordone \& Passeri, 2012). It means that it is impossible to reproduce the results reported by RLS-TT on the inconsistent instances. Due to these reasons, we mainly compare TPS with the latest and best heuristic QMST-TS on this group of benchmarks.

To ensure that the computation time required by TPS is comparable to that of QMST-TS, we independently apply our TPS algorithm 10 times to each instance, each run continues until the best found solution can not be further improved after 10 consecutive rounds of \textsl{Diversified\_Perturb} followed by \textsl{Descent\_Neighborhood\_Search} and \textsl{Explore\_Local\_Optima} search phases, or up to 50 such rounds have been applied. Experimental results show that, for all these 480 instances, TPS finds solutions no worse than the previous best known solutions\footnote{All the previous best known results of the OP1 instances could be downloaded from \url{http://www.dti.unimi.it/cordone/research/qmst.html}.}, with an accumulated CPU time ranging from less than one second to about 15 seconds, which is competitive to the runtime required by RLS-TT and QMST-TS. Specifically, for four largest instances (with $n$=50) of subclass SYM, TPS succeeds in finding improved solutions missed by previous heuristics. The results for these four instances are detailed in Table \ref{TblResultSYM}, where the first three columns indicate the instances, and the next two columns 'Best', 't(s)' respectively report the best objective value and the CPU time (in seconds) corresponding to QMST-TS, and the last four columns indicate the same information corresponding to TPS, as well as the times that TPS improves (column $\surd$) or matches (column $=$) the previous best known result among 10 runs.

\renewcommand{\baselinestretch}{0.75}\scriptsize\scriptsize
\begin{table}\centering
\begin{scriptsize}
\caption{Four improved results of the SYM subclass of group OP1}
\label{TblResultSYM}
\begin{tabular}{p{0.5cm}p{0cm}p{0.5cm}p{0cm}p{1cm}p{0cm}p{1cm}p{1cm}p{0cm}
p{1cm}p{1cm}p{0.5cm}p{0.5cm}p{0cm}}
\hline
\multicolumn{1}{c}{n} &&\multicolumn{1}{c}{m} &&\multicolumn{1}{c}{Instance Index} &&\multicolumn{2}{c}{QMST-TS} &&\multicolumn{4}{c}{TPS} &\\
\cline{7-8} \cline{10-13}

&&&&&&\centering{$Best$} &\centering{$t(s)$} &&\centering{$Best$} &\centering{$t(s)$} &\centering{$\surd$} &\centering{$=$} &\\
\hline
\centering{$50$}&&\centering{$1225$}&&\centering{$2$} &&\centering{$17600$} &\centering{$25.38$} &&\centering{$\textbf{17587}$} &\centering{$15.61$} &\centering{$5$} &\centering{$0$}  &\\
\centering{$50$}&&\centering{$1225$}&&\centering{$7$} &&\centering{$17643$} &\centering{$25.29$} &&\centering{$\textbf{17633}$} &\centering{$14.31$}  &\centering{$4$} &\centering{$0$}&\\
\centering{$50$}&&\centering{$1225$}&&\centering{$8$} &&\centering{$17685$} &\centering{$25.11$} &&\centering{$\textbf{17663}$} &\centering{$15.55$}  &\centering{$5$} &\centering{$0$}&\\
\centering{$50$}&&\centering{$1225$}&&\centering{$10$}&&\centering{$17639$} &\centering{$25.35$} &&\centering{$\textbf{17623}$} &\centering{$14.55$}  &\centering{$6$} &\centering{$0$}&\\
\hline
\end{tabular}
\end{scriptsize}
\end{table}
\renewcommand{\baselinestretch}{1.0}\small\normalsize

\subsection{Results of the SCA instances}
\label{SubsecResultsSCA}

These 6 instances are generated in by Soak, Corne, and Ahn (2006) and have been widely used as benchmarks by various heuristics, including the genetic algorithm EWD (which was originally denoted by EWD+ANX with dK-TCR), the randomized local search algorithm with tabu thresholding RLS-TT in (\"{O}ncan \& Punnen, 2010), the artificial bee colony approach ABC in (Sundar \& Singh, 2010), and the tabu search procedure QMST-TS in (Cordone \& Passeri, 2012). According to the literature, QMST-TS and ABC clearly dominate EWD and RLS-TT.

Like QMST-TS, for each one of these instances, we independently run TPS 20 times, each run using the same terminal criterion as for group OP1. The obtained results are illustrated in Table \ref{TblResultSCA}, with respect to the results reported by previous heuristics. In Table \ref{TblResultSCA}, the first column provides the problem size $n$, while the following eight columns indicate the best found results and the CPU times (in seconds) of each competing algorithm, and the last four columns show the information corresponding to our TPS algorithm, with meanings being similar to the last four columns of Table \ref{TblResultSYM}.

As shown in Table \ref{TblResultSCA}, for all these 6 instances, TPS repeatedly improves or matches the previous best results within a short time. Most importantly, for instance with $n$=80, TPS can repeatedly (12 times out of the 20 independent runs) improve the best known result. Furthermore, TPS improves 6, 6, 2, 3 results compared to EWD, RLS-TT, ABC, QMST-TS, respectively. Note that the computational time remains reasonable for each instance, being about half less than the previous fastest algorithm QMST-TS, clearly indicating the competitiveness of TPS for solving these well-known instances.

\renewcommand{\baselinestretch}{0.75}\scriptsize\scriptsize
\begin{table}[h]\centering
\begin{scriptsize}
\caption{Results of the SCA instances}
\label{TblResultSCA}
\begin{tabular}{p{0.4cm}p{0cm}p{0.5cm}p{0.5cm}p{0cm}p{0.5cm}p{0.5cm}p{0cm}p{0.5cm}p{0.5cm}p{0cm}
p{0.5cm}p{0.5cm}p{0cm}p{0.7cm}p{0.7cm}p{0.5cm}p{0.5cm}p{0cm}}
\hline
\multicolumn{1}{c}{n}& &\multicolumn{2}{c}{EWD} &&\multicolumn{2}{c}{RLS-TT} &&\multicolumn{2}{c}{ABC} &&\multicolumn{2}{c}{QMST-TS} &&\multicolumn{4}{c}{TPS} &\\
\cline{3-4} \cline{6-7} \cline{9-10} \cline{12-13} \cline{15-18}

&&\centering{$Best$} &\centering{$t(s)$} &&\centering{$Best$} &\centering{$t(s)$} &&\centering{$Best$} &\centering{$t(s)$} &&\centering{$Best$} &\centering{$t(s)$} &&\centering{$Best$} &\centering{$t(s)$} &\centering{$\surd$} &\centering{$=$} & \\
\hline
\centering{$50$} &&\centering{$25339$} &\centering{$343.0$} &&\centering{$25226$} &\centering{$3242.1$} &&\centering{$\textbf{25200}$} &\centering{$87.0$} &&\centering{$\textbf{25200}$} &\centering{$44.5$} &&\centering{$\emph{25200}$} &\centering{$15.9$} &\centering{$0$} &\centering{$20$} &\\
\centering{$60$} &&\centering{$36086$} &\centering{$495.7$} &&\centering{$35754$} &\centering{$4321.4$} &&\centering{$35466$} &\centering{$169.0$} &&\centering{$\textbf{35447}$} &\centering{$83.2$} &&\centering{$\emph{35447}$} &\centering{$43.6$} &\centering{$0$} &\centering{$20$} &\\
\centering{$70$} &&\centering{$48538$} &\centering{$716.6$} &&\centering{$48536$} &\centering{$5738.5$} &&\centering{$\textbf{48125}$} &\centering{$337.2$} &&\centering{$\textbf{48125}$} &\centering{$178.5$} &&\centering{$\emph{48125}$} &\centering{$68.3$} &\centering{$0$} &\centering{$17$} &\\
\centering{$80$} &&\centering{$63546$} &\centering{$1086.7$} &&\centering{$63546$} &\centering{$7026.3$} &&\centering{$63022$} &\centering{$417.8$} &&\centering{$63004$} &\centering{$340.4$} &&\centering{$\textbf{62963}$} &\centering{$129.6$} &\centering{$12$} &\centering{$1$} &\\
\centering{$90$} &&\centering{$79627$} &\centering{$1337.2$} &&\centering{$79922$} &\centering{$8623.6$} &&\centering{$\textbf{78879}$} &\centering{$751.8$} &&\centering{$78912$} &\centering{$579.7$} &&\centering{$\emph{78879}$} &\centering{$191.7$} &\centering{$0$} &\centering{$7$} &\\
\centering{$100$} &&\centering{$98342$} &\centering{$1828.9$} &&\centering{$98811$} &\centering{$10431.3$} &&\centering{$\textbf{96750}$} &\centering{$1542.4$} &&\centering{$96757$} &\centering{$789.4$} &&\centering{$\emph{96750}$} &\centering{$452.1$} &\centering{$0$} &\centering{$9$} &\\

\hline

\end{tabular}
\end{scriptsize}
\end{table}
\renewcommand{\baselinestretch}{1.0}\small\normalsize

\subsection{Results of the SS instances}
\label{SubsecResultsSS}

These 18 instances have been used as benchmarks by ABC (Sundar \& Singh, 2010) and QMST-TS (Cordone \& Passeri, 2012), clearly indicating that QMST-TS performs better than ABC. For comparison, for each of these instances, we independently run TPS 20 times (like QMST-TS), each run lasts until the best found solution can not be further improved after 5 consecutive rounds of \textsl{Diversified\_Perturb} followed by \textsl{Descent\_Neighborhood\_Search} and \textsl{Explore\_Local\_Optima} search phases, or up to 20 such rounds have been applied. The obtained results are listed in Table \ref{TblResultSS}, where the first two columns identify each instance, the following four columns respectively indicate the best results and the total computational time in seconds corresponding to ABC and QMST-TS, and the last four columns report the information for our TPS algorithm like in Table \ref{TblResultSCA}.

From Table \ref{TblResultSS}, one observes that for all the 12 instances with $n\ge$100, TPS can repeatedly find improved results over the best known results, while for the left 6 smaller instances with $n\le$50, TPS can easily reach the best known results. On the other hand, for every instance, TPS consumes much less computation time compared to both ABC and QMST-TS. Note that the clock frequency of our processor (2.83 GHz) is not very different from the ones corresponding to ABC (3.0 GHz) and QMST-TS (2.6 GHz), thus one can conclude that TPS is quite competitive for solving this group of instances, especially the large ones.

\renewcommand{\baselinestretch}{0.75}\scriptsize\scriptsize
\begin{table}\centering
\begin{scriptsize}
\caption{Results of the SS instances}
\label{TblResultSS}
\begin{tabular}{p{0.7cm}p{0cm}p{0.7cm}p{0cm}p{0.7cm}p{0.7cm}p{0cm}p{0.7cm}p{0.7cm}p{0cm}
p{1cm}p{1cm}p{0.4cm}p{0.4cm}p{0cm}}
\hline
\multicolumn{1}{c}{n}& &\multicolumn{1}{c}{Instance} &&\multicolumn{2}{c}{ABC} &&\multicolumn{2}{c}{QMST-TS} &&\multicolumn{4}{c}{TPS} &\\
\cline{5-6} \cline{8-9} \cline{11-14}

&&&&\centering{$Best$} &\centering{$t(s)$} & &\centering{$Best$} &\centering{$t(s)$} & &\centering{$Best$} &\centering{$t(s)$} &\centering{$\surd$} &\centering{$=$} &\\
\hline

\centering{$25$} &&\centering{$1$} &&\centering{$\textbf{5085}$} &\centering{$18.2$} &&\centering{$\textbf{5085}$} &\centering{$6.7$} &&\centering{$\emph{5085}$} &\centering{$0.9$} &\centering{$0$} &\centering{$20$} &\\
\centering{$25$} &&\centering{$2$} &&\centering{$\textbf{5081}$} &\centering{$20.4$} &&\centering{$\textbf{5081}$} &\centering{$6.6$} &&\centering{$\emph{5081}$} &\centering{$1.1$} &\centering{$0$} &\centering{$19$} &\\
\centering{$25$} &&\centering{$3$} &&\centering{$\textbf{4962}$} &\centering{$21.0$} &&\centering{$\textbf{4962}$} &\centering{$6.9$} &&\centering{$\emph{4962}$} &\centering{$0.9$} &\centering{$0$} &\centering{$20$} &\\
\centering{$50$} &&\centering{$1$} &&\centering{$\textbf{21126}$} &\centering{$173.6$} &&\centering{$\textbf{21126}$} &\centering{$50.4$} &&\centering{$\emph{21126}$} &\centering{$10.7$} &\centering{$0$} &\centering{$11$} &\\
\centering{$50$} &&\centering{$2$} &&\centering{$21123$} &\centering{$176.8$} &&\centering{$\textbf{21106}$} &\centering{$50.4$} &&\centering{$\emph{21106}$} &\centering{$11.9$} &\centering{$0$} &\centering{$7$} &\\
\centering{$50$} &&\centering{$3$} &&\centering{$\textbf{21059}$} &\centering{$190.2$} &&\centering{$\textbf{21059}$} &\centering{$50.6$} &&\centering{$\emph{21059}$} &\centering{$11.4$} &\centering{$0$} &\centering{$18$} &\\
\centering{$100$} &&\centering{$1$} &&\centering{$89098$} &\centering{$2333.2$} &&\centering{$88871$} &\centering{$965.8$} &&\centering{$\textbf{88790}$} &\centering{$216.9$} &\centering{$2$} &\centering{$0$} &\\
\centering{$100$} &&\centering{$2$} &&\centering{$89202$} &\centering{$2319.0$} &&\centering{$89049$} &\centering{$957.7$} &&\centering{$\textbf{88851}$} &\centering{$212.4$} &\centering{$5$} &\centering{$0$} &\\
\centering{$100$} &&\centering{$3$} &&\centering{$89007$} &\centering{$1977.6$} &&\centering{$88720$} &\centering{$961.2$} &&\centering{$\textbf{88631}$} &\centering{$232.3$} &\centering{$2$} &\centering{$0$} &\\
\centering{$150$} &&\centering{$1$} &&\centering{$205619$} &\centering{$8897.4$} &&\centering{$205615$} &\centering{$2928.7$} &&\centering{$\textbf{205201}$} &\centering{$847.1$} &\centering{$4$} &\centering{$0$} &\\
\centering{$150$} &&\centering{$2$} &&\centering{$205874$} &\centering{$7486.6$} &&\centering{$205509$} &\centering{$2923.0$} &&\centering{$\textbf{205013}$} &\centering{$835.6$} &\centering{$3$} &\centering{$0$} &\\
\centering{$150$} &&\centering{$3$} &&\centering{$205634$} &\centering{$8658.6$} &&\centering{$205094$} &\centering{$2928.6$} &&\centering{$\textbf{204841}$} &\centering{$880.6$} &\centering{$1$} &\centering{$0$} &\\
\centering{$200$} &&\centering{$1$} &&\centering{$371797$} &\centering{$22828.4$} &&\centering{$371492$} &\centering{$6320.3$} &&\centering{$\textbf{371104}$} &\centering{$2206.1$} &\centering{$3$} &\centering{$0$} &\\
\centering{$200$} &&\centering{$2$} &&\centering{$371864$} &\centering{$23112.0$} &&\centering{$371698$} &\centering{$6332.1$} &&\centering{$\textbf{370916}$} &\centering{$2397.1$} &\centering{$14$} &\centering{$0$} &\\
\centering{$200$} &&\centering{$3$} &&\centering{$372156$} &\centering{$25534.2$} &&\centering{$371584$} &\centering{$6324.3$} &&\centering{$\textbf{371081}$} &\centering{$2742.7$} &\centering{$9$} &\centering{$0$} &\\
\centering{$250$} &&\centering{$1$} &&\centering{$587924$} &\centering{$51268.2$} &&\centering{$586861$} &\centering{$9572.3$} &&\centering{$\textbf{586445}$} &\centering{$5212.4$} &\centering{$1$} &\centering{$0$} &\\
\centering{$250$} &&\centering{$2$} &&\centering{$588068$} &\centering{$56818.2$} &&\centering{$587607$} &\centering{$9592.9$} &&\centering{$\textbf{586500}$} &\centering{$4461.7$} &\centering{$14$} &\centering{$0$} &\\
\centering{$250$} &&\centering{$3$} &&\centering{$587883$} &\centering{$46565.8$} &&\centering{$587281$} &\centering{$9601.2$} &&\centering{$\textbf{586964}$} &\centering{$4701.5$} &\centering{$6$} &\centering{$0$} &\\

\hline
\end{tabular}
\end{scriptsize}
\end{table}
\renewcommand{\baselinestretch}{1.0}\small\normalsize

\subsection{Results of the RAND and SOAK instances}
\label{SubsecResultsRand}

\renewcommand{\baselinestretch}{0.75}\scriptsize\scriptsize
\begin{table}\centering
\begin{scriptsize}
\caption{Results of the RAND instances}
\label{TblResultRAND}
\begin{tabular}{p{2cm}p{0.7cm}p{0.7cm}p{0cm}p{0.7cm}p{0.7cm}p{0cm}p{0.7cm}p{0.7cm}p{0cm}p{0.7cm}p{0.7cm}p{0.2cm}p{0.2cm}p{0cm}}
\hline
Instance &\multicolumn{2}{c}{ABC} &&\multicolumn{2}{c}{ITS} &&\multicolumn{2}{c}{HSII} &&\multicolumn{2}{c}{TPS} &\\
\cline{2-3} \cline{5-6} \cline{8-9} \cline{11-14}

&\centering{$Best$} &\centering{$t(s)$} &&\centering{$Best$} &\centering{$t(s)$} &&\centering{$Best$} &\centering{$t(s)$} &&\centering{$Best$} &\centering{$t(s)$} &\centering{$\surd$} &\centering{$=$}&\\
\hline

\centering{RAND-150-1} &\centering{$194294$} &\centering{$4000$} &&\centering{$192946$} &\centering{$4000$} &&\centering{$192606$} &\centering{$4000$} &&\centering{$\textbf{192296}$} &\centering{$4000$} &\centering{$2$} &\centering{$0$} &\\
\centering{RAND-150-2} &\centering{$194218$} &\centering{$4000$} &&\centering{$193034$} &\centering{$4000$} &&\centering{$192607$} &\centering{$4000$} &&\centering{$\textbf{192597}$} &\centering{$4000$} &\centering{$1$} &\centering{$0$} &\\
\centering{RAND-150-3} &\centering{$193882$} &\centering{$4000$} &&\centering{$192965$} &\centering{$4000$} &&\centering{$192577$} &\centering{$4000$} &&\centering{$\textbf{192439}$} &\centering{$4000$} &\centering{$1$} &\centering{$0$} &\\
\centering{RAND-200-1} &\centering{$353163$} &\centering{$12000$} &&\centering{$351216$} &\centering{$12000$} &&\centering{$350517$} &\centering{$12000$} &&\centering{$\textbf{350294}$} &\centering{$12000$} &\centering{$1$} &\centering{$0$} &\\
\centering{RAND-200-2} &\centering{$353784$} &\centering{$12000$} &&\centering{$351312$} &\centering{$12000$} &&\centering{$\textbf{350389}$} &\centering{$12000$} &&\centering{$350995$} &\centering{$12000$} &\centering{$0$} &\centering{$0$} &\\
\centering{RAND-200-3} &\centering{$353169$} &\centering{$12000$} &&\centering{$351466$} &\centering{$12000$} &&\centering{$351057$} &\centering{$12000$} &&\centering{$\textbf{350929}$} &\centering{$12000$} &\centering{$2$} &\centering{$0$} &\\
\centering{RAND-250-1} &\centering{$561864$} &\centering{$20000$} &&\centering{$558451$} &\centering{$20000$} &&\centering{$\textbf{556929}$} &\centering{$20000$} &&\centering{$557649$} &\centering{$20000$} &\centering{$0$} &\centering{$0$} &\\
\centering{RAND-250-2} &\centering{$560704$} &\centering{$20000$} &&\centering{$558820$} &\centering{$20000$} &&\centering{$557474$} &\centering{$20000$} &&\centering{$\textbf{557044}$} &\centering{$20000$} &\centering{$1$} &\centering{$0$} &\\
\centering{RAND-250-3} &\centering{$561497$} &\centering{$20000$} &&\centering{$559304$} &\centering{$20000$} &&\centering{$556813$} &\centering{$20000$} &&\centering{$\textbf{556806}$} &\centering{$20000$} &\centering{$1$} &\centering{$0$} &\\

\hline
\end{tabular}
\end{scriptsize}
\end{table}
\renewcommand{\baselinestretch}{1.0}\small\normalsize

\renewcommand{\baselinestretch}{0.75}\scriptsize\scriptsize
\begin{table}\centering
\begin{scriptsize}
\caption{Results of the SOAK instances}
\label{TblResultSOAK}
\begin{tabular}{p{2cm}p{0.7cm}p{0.7cm}p{0cm}p{0.7cm}p{0.7cm}p{0cm}p{0.7cm}p{0.7cm}p{0cm}p{0.7cm}p{0.7cm}p{0.2cm}p{0.2cm}p{0cm}}
\hline
Instance &\multicolumn{2}{c}{ABC} &&\multicolumn{2}{c}{ITS} &&\multicolumn{2}{c}{HSII} &&\multicolumn{2}{c}{TPS} &\\
\cline{2-3} \cline{5-6} \cline{8-9} \cline{11-14}

&\centering{$Best$} &\centering{$t(s)$} &&\centering{$Best$} &\centering{$t(s)$} &&\centering{$Best$} &\centering{$t(s)$} &&\centering{$Best$} &\centering{$t(s)$} &\centering{$\surd$} &\centering{$=$}&\\
\hline

\centering{SOAK-150-1} &\centering{$207652$} &\centering{$4000$} &&\centering{$\textbf{206721}$} &\centering{$4000$} &&\centering{$206925$} &\centering{$4000$} &&\centering{$\emph{206721}$} &\centering{$4000$} &\centering{$0$} &\centering{$2$} &\\
\centering{SOAK-150-2} &\centering{$208206$} &\centering{$4000$} &&\centering{$\textbf{206761}$} &\centering{$4000$} &&\centering{$207102$} &\centering{$4000$} &&\centering{$\emph{206761}$} &\centering{$4000$} &\centering{$0$} &\centering{$3$} &\\
\centering{SOAK-150-3} &\centering{$207533$} &\centering{$4000$} &&\centering{$206802$} &\centering{$4000$} &&\centering{$206781$} &\centering{$4000$} &&\centering{$\textbf{206759}$} &\centering{$4000$} &\centering{$1$} &\centering{$1$} &\\
\centering{SOAK-200-1} &\centering{$372419$} &\centering{$12000$} &&\centering{$370137$} &\centering{$12000$} &&\centering{$370265$} &\centering{$12000$} &&\centering{$\textbf{369807}$} &\centering{$12000$} &\centering{$2$} &\centering{$0$} &\\
\centering{SOAK-200-2} &\centering{$371641$} &\centering{$12000$} &&\centering{$370028$} &\centering{$12000$} &&\centering{$369982$} &\centering{$12000$} &&\centering{$\textbf{369878}$} &\centering{$12000$} &\centering{$1$} &\centering{$0$} &\\
\centering{SOAK-200-3} &\centering{$372117$} &\centering{$12000$} &&\centering{$370046$} &\centering{$12000$} &&\centering{$370045$} &\centering{$12000$} &&\centering{$\textbf{369775}$} &\centering{$12000$} &\centering{$3$} &\centering{$0$} &\\
\centering{SOAK-250-1} &\centering{$584799$} &\centering{$20000$} &&\centering{$582282$} &\centering{$20000$} &&\centering{$581819$} &\centering{$20000$} &&\centering{$\textbf{581536}$} &\centering{$20000$} &\centering{$1$} &\centering{$0$} &\\
\centering{SOAK-250-2} &\centering{$584409$} &\centering{$20000$} &&\centering{$582145$} &\centering{$20000$} &&\centering{$581691$} &\centering{$20000$} &&\centering{$\textbf{581438}$} &\centering{$20000$} &\centering{$3$} &\centering{$0$} &\\
\centering{SOAK-250-3} &\centering{$585717$} &\centering{$20000$} &&\centering{$582708$} &\centering{$20000$} &&\centering{$581854$} &\centering{$20000$} &&\centering{$\textbf{581819}$} &\centering{$20000$} &\centering{$1$} &\centering{$0$} &\\

\hline
\end{tabular}
\end{scriptsize}
\end{table}
\renewcommand{\baselinestretch}{1.0}\small\normalsize

Very recently, Lozano et al. (2013) propose a hybrid heuristic named HSII and evaluate its performance using two groups (RAND and SOAK) of 18 newly generated benchmarks, in comparison with two previous heuristics, i.e., ITS (Palubeckis, Rubliauskas, \& Targamadz\`{e}, 2010) and ABC (Sundar \& Singh, 2010)\footnote{ITS and ABC did not report their results on groups RAND, SOAK, and QAP-QMSTP. In order to reproduce the results corresponding to ITS and ABC on these benchmarks, Lozano et al. (2013) download the source code of ITS from \url{http://www.soften.ktu.lt/~gintaras/qmstp.html} and re-implement the ABC algorithm, and then compare the obtained results with respect to their HSII algorithm (tested on the same platform).}. For each instance, the above three algorithms are respectively executed 10 independent times, each run stops using a time limit that varies according to the problem size (400, 1200, 2000 seconds for instances with $n$=150, 200, 250, respectively on a 3.2 GHz Intel processor with 12 GB RAM). To evaluate our TPS algorithm under a comparable condition, we also independently run TPS 10 times to solve each instance, using the same cutoff time like in (Lozano et al., 2013) for each run (we use a computer with an Intel Xeon E5440 2.83 GHz processor and 2 GB RAM). The obtained results are provided in Tables \ref{TblResultRAND} and \ref{TblResultSOAK}, where the meanings of the columns are similar to those in previous Tables.

On one hand, as listed in Table \ref{TblResultRAND}, for seven out of the nine instances of group RAND, TPS succeeds in finding an improved solution over the compared algorithms, while for the left two instances (\textsl{Rand-200-2}, \textsl{Rand-250-1}), TPS fails to match the previous best known results within the limited runtime. On the other hand, for the nine SOAK instances (Table \ref{TblResultSOAK}), TPS improves seven best known results and matches the left two results. Since all the algorithms are executed under the same cutoff time, and the clock frequency of our processor (2.83 GHz) is lower than the processor (3.2 GHz) used to test ITS, ABC and HSII (Lozano et al., 2013), one can conclude that TPS is competitive for solving these two groups of large instances, with respect to the reference heuristics.

\subsection{Results of the QAP-QMSTP instances}
\label{SubsecResultsOP2}

This group of 29 special QMSTP instances are transformed from the \textsl{quadratic assignment problem}, including 14 CHR ones (Christofides \& Benavent, 1989) and 15 NUG ones (Nugent, Vollman, \& Ruml, 1968), while guaranteeing an one-to-one correspondence of the feasible solutions after transformation (\"{O}ncan \& Punnen, 2010). Note that, due to the rather peculiar problem structures of these transformed instances, many of them are extremely difficult for the existing QMSTP algorithms to reach optimality, although the original QAP instances have all been solved to optimality by QAP algorithms (Burkard, Karisch, \& Rendl, 1997).

As mentioned in Section \ref{SubsecParameter}, for these transformed instances, we reset parameter $p$ to 0.5 instead of 1 as for the other groups of benchmarks (the extra tabu tenure $l_{swap}$ used in the swap-vertex based directed perturbation operator is distributed at random within $[n,5n]$), and enlarge parameter $L_{dir}$ to belong to $[5n,10n]$, while keeping all the other ingredients and parameters unchanged. For each of these 29 instances, we independently run TPS 10 times, each run continues until the best found solution can not be further improved after 100 consecutive rounds of \textsl{Diversified\_Perturb} followed by \textsl{Descent\_Neighborhood\_Search} and \textsl{Explore\_Local\_Optima}, to ensure that the accumulated runtime remains comparable with respect to the compared heuristics.

Table \ref{TblResultOP2} lists in detail the obtained results. Respectively, the first column reports the instance name, and the second one its optimal value known from the QAP literature (Burkard, Karisch, \& Rendl, 1997)\footnote{The optimal solutions of the original QAP instances are available online at the QAPLIB: \url{http://www.seas.upenn.edu/qaplib}.}. The following 10 columns report the best found results and the accumulated CPU times (in seconds) of each compared algorithm, i.e., RLS-TT (\"{O}ncan \& Punnen, 2010), ABC (Sundar \& Singh, 2010), ITS (Palubeckis, Rubliauskas, \& Targamadz\`{e}, 2010), HSII (Lozano et al., 2013) and QMST-TS with re-tuned parameters (Cordone \& Passeri, 2012). Note that, like for the RAND and SOAK instances, the results corresponding to ABC and ITS are reproduced by Lozano et al. (2013). The final three columns indicate the best found results and the accumulated CPU time corresponding to our TPS algorithm, as well as the times that TPS hits the optimal solution among the 10 independent runs.

\renewcommand{\baselinestretch}{0.75}\scriptsize\scriptsize
\begin{table}\centering
\begin{scriptsize}
\caption{Results of $TPS$ on the QAP-QMSTP instances compared with RLS-TT (\"{O}ncan \& Punnen, 2010), ABC (Sundar \& Singh, 2010), ITS (Palubeckis, Rubliauskas, \& Targamadz\`{e}, 2010), HSII (Lozano et al., 2013) and QMST-TS with re-tuned parameters (Cordone \& Passeri, 2012).}
\label{TblResultOP2}
\begin{tabular}{p{0.4cm}p{0cm}p{0.4cm}p{0cm}p{0.4cm}p{0.4cm}p{0cm}p{0.4cm}p{0.4cm}p{0cm}p{0.4cm}p{0.4cm}p{0cm}p{0.4cm}p{0.4cm}p{0cm}p{0.4cm}p{0.4cm}p{0cm}
p{0.5cm}p{0.4cm}p{0.4cm}p{0cm}}
\hline
Instance& &Opt. &&\multicolumn{2}{c}{RLS-TT} &&\multicolumn{2}{c}{ABC} &&\multicolumn{2}{c}{ITS} &&\multicolumn{2}{c}{HSII} &&\multicolumn{2}{c}{QMST-TS} &&\multicolumn{3}{c}{TPS} &\\
\cline{5-6} \cline{8-9} \cline{11-12} \cline{14-15} \cline{17-18} \cline{20-22}

&&&&\centering{$Best$} &\centering{$t(s)$} &&\centering{$Best$} &\centering{$t(s)$} &&\centering{$Best$} &\centering{$t(s)$} &&\centering{$Best$} &\centering{$t(s)$} &&\centering{$Best$} &\centering{$t(s)$} &&\centering{$Best$} &\centering{$t(s)$} &\centering{$Hit$} & \\
\hline

\hline

\centering{$chr12a$} &&\centering{$9552$} &&\centering{$11170$} &\centering{$783$} &&\centering{$14290$} &\centering{$10000$} &&\centering{$16694$} &\centering{$10000$} &&\centering{$\emph{9552}$} &\centering{$10000$} &&\centering{$\emph{9552}$} &\centering{$288$} &&\centering{$\emph{9552}$} &\centering{$62$}&\centering{$10$} &\\
\centering{$chr12b$} &&\centering{$9742$} &&\centering{$10753$} &\centering{$790$} &&\centering{$21552$} &\centering{$10000$} &&\centering{$16356$} &\centering{$10000$} &&\centering{$\emph{9742}$} &\centering{$10000$} &&\centering{$\emph{9742}$} &\centering{$287$} &&\centering{$\emph{9742}$} &\centering{$67$}&\centering{$10$} &\\
\centering{$chr12c$} &&\centering{$11156$} &&\centering{$12712$} &\centering{$783$} &&\centering{$15810$} &\centering{$10000$} &&\centering{$17434$} &\centering{$10000$} &&\centering{$\emph{11156}$} &\centering{$10000$} &&\centering{$\emph{11156}$} &\centering{$286$} &&\centering{$\emph{11156}$} &\centering{$84$}&\centering{$10$} &\\
\centering{$chr15a$} &&\centering{$9896$} &&\centering{$11638$} &\centering{$1239$} &&\centering{$24224$} &\centering{$10000$} &&\centering{$16718$} &\centering{$10000$} &&\centering{$9952$} &\centering{$10000$} &&\centering{$9936$} &\centering{$497$} &&\centering{$\textbf{9896}$} &\centering{$168$}&\centering{$10$} &\\
\centering{$chr15b$} &&\centering{$7990$} &&\centering{$10145$} &\centering{$1136$} &&\centering{$28340$} &\centering{$10000$} &&\centering{$17208$} &\centering{$10000$} &&\centering{$8384$} &\centering{$10000$} &&\centering{$\emph{7990}$} &\centering{$492$} &&\centering{$\emph{7990}$} &\centering{$158$}&\centering{$10$} &\\
\centering{$chr15c$} &&\centering{$9504$} &&\centering{$12769$} &\centering{$1254$} &&\centering{$25566$} &\centering{$10000$} &&\centering{$19302$} &\centering{$10000$} &&\centering{$\emph{9504}$} &\centering{$10000$} &&\centering{$\emph{9504}$} &\centering{$492$} &&\centering{$\emph{9504}$} &\centering{$147$}&\centering{$10$} &\\
\centering{$chr18a$} &&\centering{$11098$} &&\centering{$12757$} &\centering{$3325$} &&\centering{$24954$} &\centering{$10000$} &&\centering{$22496$} &\centering{$10000$} &&\centering{$13834$} &\centering{$10000$} &&\centering{$\emph{11098}$} &\centering{$793$} &&\centering{$\emph{11098}$} &\centering{$282$}&\centering{$9$} &\\
\centering{$chr18b$} &&\centering{$1534$} &&\centering{$1676$} &\centering{$3354$} &&\centering{$2160$} &\centering{$10000$} &&\centering{$\emph{1534}$} &\centering{$10000$} &&\centering{$\emph{1534}$} &\centering{$10000$} &&\centering{$\emph{1534}$} &\centering{$789$} &&\centering{$\emph{1534}$} &\centering{$209$}&\centering{$10$} &\\
\centering{$chr20a$} &&\centering{$2192$} &&\centering{$2445$} &\centering{$4968$} &&\centering{$4742$} &\centering{$10000$} &&\centering{$2232$} &\centering{$10000$} &&\centering{$2276$} &\centering{$10000$} &&\centering{$\emph{2192}$} &\centering{$1043$} &&\centering{$\emph{2192}$} &\centering{$544$}&\centering{$3$} &\\
\centering{$chr20b$} &&\centering{$2298$} &&\centering{$2730$} &\centering{$4652$} &&\centering{$3704$} &\centering{$10000$} &&\centering{$2440$} &\centering{$10000$} &&\centering{$2462$} &\centering{$10000$} &&\centering{$2352$} &\centering{$1044$} &&\centering{$\textbf{2298}$} &\centering{$561$}&\centering{$1$} &\\
\centering{$chr20c$} &&\centering{$14142$} &&\centering{$30124$} &\centering{$4763$} &&\centering{$49842$} &\centering{$10000$} &&\centering{$36558$} &\centering{$10000$} &&\centering{$20206$} &\centering{$10000$} &&\centering{$14202$} &\centering{$1046$} &&\centering{$\textbf{14142}$} &\centering{$395$}&\centering{$10$} &\\
\centering{$chr22a$} &&\centering{$6156$} &&\centering{$8760$} &\centering{$5089$} &&\centering{$8688$} &\centering{$10000$} &&\centering{$6390$} &\centering{$10000$} &&\centering{$6334$} &\centering{$10000$} &&\centering{$6228$} &\centering{$1395$} &&\centering{$\textbf{6156}$} &\centering{$702$}&\centering{$9$} &\\
\centering{$chr22b$} &&\centering{$6194$} &&\centering{$8402$} &\centering{$4741$} &&\centering{$8908$} &\centering{$10000$} &&\centering{$6314$} &\centering{$10000$} &&\centering{$6396$} &\centering{$10000$} &&\centering{$6314$} &\centering{$1422$} &&\centering{$\textbf{6194}$} &\centering{$953$}&\centering{$1$} &\\
\centering{$chr25a$} &&\centering{$3796$} &&\centering{$9658$} &\centering{$5223$} &&\centering{$8540$} &\centering{$10000$} &&\centering{$4300$} &\centering{$10000$} &&\centering{$4310$} &\centering{$10000$} &&\centering{$3866$} &\centering{$2135$} &&\centering{$\textbf{3796}$} &\centering{$1484$}&\centering{$3$} &\\

\hline

\centering{$nug12$} &&\centering{$578$} &&\centering{$605$} &\centering{$639$} &&\centering{$656$} &\centering{$10000$} &&\centering{$\emph{578}$} &\centering{$10000$} &&\centering{$\emph{578}$} &\centering{$10000$} &&\centering{$\emph{578}$} &\centering{$287$} &&\centering{$\emph{578}$} &\centering{$61$}&\centering{$10$} &\\
\centering{$nug14$} &&\centering{$1014$} &&\centering{$1084$} &\centering{$724$} &&\centering{$1140$} &\centering{$10000$} &&\centering{$\emph{1014}$} &\centering{$10000$} &&\centering{$1026$} &\centering{$10000$} &&\centering{$\emph{1014}$} &\centering{$414$} &&\centering{$\emph{1014}$} &\centering{$101$}&\centering{$10$} &\\
\centering{$nug15$} &&\centering{$1150$} &&\centering{$1265$} &\centering{$1348$} &&\centering{$1404$} &\centering{$10000$} &&\centering{$\emph{1150}$} &\centering{$10000$} &&\centering{$1152$} &\centering{$10000$} &&\centering{$\emph{1150}$} &\centering{$493$} &&\centering{$\emph{1150}$} &\centering{$122$}&\centering{$10$} &\\
\centering{$nug16a$} &&\centering{$1610$} &&\centering{$1742$} &\centering{$2311$} &&\centering{$1944$} &\centering{$10000$} &&\centering{$1638$} &\centering{$10000$} &&\centering{$1634$} &\centering{$10000$} &&\centering{$1622$} &\centering{$583$} &&\centering{$\textbf{1610}$} &\centering{$163$}&\centering{$10$} &\\
\centering{$nug16b$} &&\centering{$1240$} &&\centering{$1350$} &\centering{$2936$} &&\centering{$1480$} &\centering{$10000$} &&\centering{$1248$} &\centering{$10000$} &&\centering{$1246$} &\centering{$10000$} &&\centering{$\emph{1240}$} &\centering{$579$} &&\centering{$\emph{1240}$} &\centering{$143$}&\centering{$10$} &\\
\centering{$nug17$} &&\centering{$1732$} &&\centering{$1874$} &\centering{$3422$} &&\centering{$2066$} &\centering{$10000$} &&\centering{$1768$} &\centering{$10000$} &&\centering{$1774$} &\centering{$10000$} &&\centering{$1750$} &\centering{$685$} &&\centering{$\textbf{1732}$} &\centering{$197$}&\centering{$10$} &\\
\centering{$nug18$} &&\centering{$1930$} &&\centering{$2056$} &\centering{$3482$} &&\centering{$2224$} &\centering{$10000$} &&\centering{$1964$} &\centering{$10000$} &&\centering{$1984$} &\centering{$10000$} &&\centering{$1942$} &\centering{$793$} &&\centering{$\textbf{1930}$} &\centering{$229$}&\centering{$10$} &\\
\centering{$nug20$} &&\centering{$2570$} &&\centering{$2860$} &\centering{$5151$} &&\centering{$2900$} &\centering{$10000$} &&\centering{$2644$} &\centering{$10000$} &&\centering{$2662$} &\centering{$10000$} &&\centering{$2580$} &\centering{$1047$} &&\centering{$\textbf{2570}$} &\centering{$340$}&\centering{$10$} &\\
\centering{$nug21$} &&\centering{$2438$} &&\centering{$2698$} &\centering{$5184$} &&\centering{$3042$} &\centering{$10000$} &&\centering{$2502$} &\centering{$10000$} &&\centering{$2540$} &\centering{$10000$} &&\centering{$2488$} &\centering{$1201$} &&\centering{$\textbf{2438}$} &\centering{$438$}&\centering{$10$} &\\
\centering{$nug22$} &&\centering{$3596$} &&\centering{$3868$} &\centering{$5482$} &&\centering{$4580$} &\centering{$10000$} &&\centering{$3712$} &\centering{$10000$} &&\centering{$3750$} &\centering{$10000$} &&\centering{$3672$} &\centering{$1371$} &&\centering{$\textbf{3596}$} &\centering{$509$}&\centering{$10$} &\\
\centering{$nug24$} &&\centering{$3488$} &&\centering{$3874$} &\centering{$5914$} &&\centering{$4340$} &\centering{$10000$} &&\centering{$3648$} &\centering{$10000$} &&\centering{$3688$} &\centering{$10000$} &&\centering{$3590$} &\centering{$1838$} &&\centering{$\textbf{3488}$} &\centering{$750$}&\centering{$10$} &\\
\centering{$nug25$} &&\centering{$3744$} &&\centering{$4083$} &\centering{$5983$} &&\centering{$4522$} &\centering{$10000$} &&\centering{$3954$} &\centering{$10000$} &&\centering{$3940$} &\centering{$10000$} &&\centering{$3874$} &\centering{$2098$} &&\centering{$\textbf{3744}$} &\centering{$933$}&\centering{$10$} &\\
\centering{$nug27$} &&\centering{$5234$} &&\centering{$5966$} &\centering{$6025$} &&\centering{$6284$} &\centering{$10000$} &&\centering{$5456$} &\centering{$10000$} &&\centering{$5534$} &\centering{$10000$} &&\centering{$5352$} &\centering{$2788$} &&\centering{$\textbf{5234}$} &\centering{$1210$}&\centering{$10$} &\\
\centering{$nug28$} &&\centering{$5166$} &&\centering{$5819$} &\centering{$6087$} &&\centering{$6238$} &\centering{$10000$} &&\centering{$5406$} &\centering{$10000$} &&\centering{$5484$} &\centering{$10000$} &&\centering{$5262$} &\centering{$3228$} &&\centering{$\textbf{5166}$} &\centering{$1707$}&\centering{$10$} &\\
\centering{$nug30$} &&\centering{$6124$} &&\centering{$6923$} &\centering{$6227$} &&\centering{$7688$} &\centering{$10000$} &&\centering{$6506$} &\centering{$10000$} &&\centering{$6528$} &\centering{$10000$} &&\centering{$6364$} &\centering{$4283$} &&\centering{$\textbf{6124}$} &\centering{$2204$}&\centering{$8$} &\\

\hline
\end{tabular}
\end{scriptsize}
\end{table}
\renewcommand{\baselinestretch}{1.0}\small\normalsize

From Table \ref{TblResultOP2}, one first observes that the previous QMSTP algorithms RLS-TT, ABC, ITS, HSII, QMST-TS respectively miss 29, 29, 25, 23, 17 optimal solutions. On the contrary, for all these instances, our TPS algorithm can consistently match the optimal solutions, while consuming statistically much less CPU time compared to the previous best and fastest algorithm, clearly indicating its competitiveness for solving these particular and challenging instances, not only in terms of solution quality, but also in terms of runtime. In the next Section, we provide insights into the importance of the swap-vertex based directed perturbation operator for solving these special instances.

\section{Discussions}
\label{SecAnalyze}

In this Section, we discuss the impact of several important ingredients on the performance of the proposed algorithm, including the fast examination criterion, the directed perturbation operators, as well as the diversified perturbation operator.

\subsection{Importance of the fast examination criterion}
\label{SubsecAnalyzeFast}

As described in Section \ref{SubsecFastEvalute}, our TPS algorithm employs a pre-estimation criterion to discard useless swap-edge moves to reduce the neighborhood to be examined at each iteration of the TPS search process. In this section, we highlight the importance of this fast examination technique.

For this, while solving each group of instances (Section \ref{SecResults}), we record the total number of all the possible edges $e \in E\backslash X$, associated with the number of the useless edges discarded by the fast examination criterion, just as detailed in Table \ref{TblResultAnalyzeFast}.

As shown in Table \ref{TblResultAnalyzeFast}, for benchmarks CP, OP1, SCA, SS, RAND, SOAK, QAP-QMSTP, the fast examination criterion can respectively identify and discard $89.6\%$, $74.2\%$, $93.4\%$, $97.0\%$, $97.2\%$, $97.1\%$, $62.8\%$ useless edges among all the possible edges needing evaluation. Clearly, this technique significantly speeds up the search process, indicating its importance to the proposed algorithm.

\renewcommand{\baselinestretch}{0.75}\scriptsize\scriptsize
\begin{table}\centering
\begin{scriptsize}
\caption{Importance of the fast evaluation criterion}
\label{TblResultAnalyzeFast}
\begin{tabular}{p{2.5cm}p{0cm}p{2.5cm}p{0cm}p{2.5cm}p{0cm}p{2.5cm}p{0cm}}
\hline
\centering{Group} &&\centering{Total} &&\centering{Discarded} &&\centering{Ratio} &\\
\hline

\centering{CP}        &&\centering{$90459569$} &&\centering{$81100273$} &&\centering{$89.6\%$} &\\
\centering{OP1}       &&\centering{$14455279$} &&\centering{$10724538$} &&\centering{$74.2\%$} &\\
\centering{SCA}       &&\centering{$31016185$} &&\centering{$28964826$} &&\centering{$93.4\%$} &\\
\centering{SS}        &&\centering{$623594497$} &&\centering{$604686287$} &&\centering{$97.0\%$} &\\
\centering{RAND}      &&\centering{$2167317367$} &&\centering{$2106191490$} &&\centering{$97.2\%$} &\\
\centering{SOAK}      &&\centering{$2319268727$} &&\centering{$2252268468$} &&\centering{$97.1\%$} &\\
\centering{QAP-QMSTP} &&\centering{$248722385$} &&\centering{$156296256$} &&\centering{$62.8\%$} &\\

\hline
\end{tabular}
\end{scriptsize}
\end{table}
\renewcommand{\baselinestretch}{1.0}\small\normalsize

\subsection{Impact of the directed perturbation operators}
\label{SubsecAnalyzeDirectedPerturb}

\renewcommand{\baselinestretch}{0.75}\scriptsize\scriptsize
\begin{table}\centering
\begin{scriptsize}
\caption{Results corresponding to different directed perturbation operators by varying parameter $p$}
\label{TblAnaylyzeDirectPerturb}
\begin{tabular}{p{2cm}p{0.8cm}p{1.0cm}p{0.8cm}p{0cm}p{0.8cm}p{1.0cm}p{0.8cm}p{0cm}p{0.8cm}p{1.0cm}p{0.8cm}p{0cm}}
\hline
Instance &\multicolumn{3}{c}{$p=1$} &&\multicolumn{3}{c}{$p=0.5$} &&\multicolumn{3}{c}{$p=0$} &\\
\cline{2-4} \cline{6-8} \cline{10-12}

&\centering{$Best$} &\centering{$Average$} &\centering{$t(s)$} &&\centering{$Best$} &\centering{$Average$} &\centering{$t(s)$} &&\centering{$Best$} &\centering{$Average$} &\centering{$t(s)$} &\\
\hline

\centering{SS-100-1} &\centering{$\textbf{88790}$} &\centering{$\textbf{89031.4}$} &\centering{$119.9$} &&\centering{$88985$} &\centering{$89156.1$} &\centering{$178.8$} &&\centering{$88876$} &\centering{$89404.2$} &\centering{$207.8$} &\\
\centering{SS-100-2} &\centering{$88941$} &\centering{$\textbf{89159.2}$} &\centering{$107.4$} &&\centering{$\textbf{88881}$} &\centering{$89217.7$} &\centering{$168.3$} &&\centering{$89240$} &\centering{$89450.8$} &\centering{$203.5$} &\\
\centering{SS-100-3} &\centering{$\textbf{88631}$} &\centering{$\textbf{88966.4}$} &\centering{$120.9$} &&\centering{$88898$} &\centering{$89139.3$} &\centering{$179.2$} &&\centering{$89114$} &\centering{$89471.1$} &\centering{$215.3$} &\\
\centering{SS-150-1} &\centering{$\textbf{205201}$} &\centering{$\textbf{205757.0}$} &\centering{$467.3$} &&\centering{$205445$} &\centering{$205868.9$} &\centering{$681.4$} &&\centering{$205818$} &\centering{$206143.2$} &\centering{$1293.5$} &\\
\centering{SS-150-2} &\centering{$\textbf{205013}$} &\centering{$\textbf{205714.6}$} &\centering{$411.5$} &&\centering{$205451$} &\centering{$205917.2$} &\centering{$706.4$} &&\centering{$206092$} &\centering{$206463.6$} &\centering{$1132.6$} &\\
\centering{SS-150-3} &\centering{$\textbf{204841}$} &\centering{$\textbf{205505.7}$} &\centering{$462.3$} &&\centering{$205518$} &\centering{$205830.0$} &\centering{$702.4$} &&\centering{$205719$} &\centering{$206288.0$} &\centering{$1088.4$} &\\
\centering{SS-200-1} &\centering{$\textbf{371218}$} &\centering{$\textbf{371827.7}$} &\centering{$1105.9$} &&\centering{$371618$} &\centering{$372180.8$} &\centering{$1820.9$} &&\centering{$372150$} &\centering{$372686.6$} &\centering{$4083.3$} &\\
\centering{SS-200-2} &\centering{$\textbf{371081}$} &\centering{$\textbf{371556.2}$} &\centering{$1317.1$} &&\centering{$371659$} &\centering{$371989.3$} &\centering{$2178.7$} &&\centering{$371952$} &\centering{$372503.4$} &\centering{$4108.4$} &\\
\centering{SS-200-3} &\centering{$\textbf{371081}$} &\centering{$\textbf{371448.8}$} &\centering{$1524.6$} &&\centering{$371653$} &\centering{$372001.4$} &\centering{$2049.5$} &&\centering{$372335$} &\centering{$372808.7$} &\centering{$4108.8$} &\\
\centering{SS-250-1} &\centering{$\textbf{586445}$} &\centering{$\textbf{587322.1}$} &\centering{$2461.6$} &&\centering{$587061$} &\centering{$587488.2$} &\centering{$6264.8$} &&\centering{$587880$} &\centering{$588552.9$} &\centering{$11426.9$} &\\
\centering{SS-250-2} &\centering{$587210$} &\centering{$\textbf{587526.1}$} &\centering{$2193.8$} &&\centering{$\textbf{586982}$} &\centering{$587575.8$} &\centering{$8632.7$} &&\centering{$587950$} &\centering{$588562.7$} &\centering{$16061.8$} &\\
\centering{SS-250-3} &\centering{$\textbf{586964}$} &\centering{$\textbf{587486.5}$} &\centering{$2472.0$} &&\centering{$587122$} &\centering{$587994.5$} &\centering{$4486.3$} &&\centering{$587752$} &\centering{$588774.0$} &\centering{$8354.9$} &\\
\\
\centering{RAND-150-1} &\centering{$\textbf{192296}$} &\centering{$\textbf{192909.4}$} &\centering{$4000.0$} &&\centering{$193123$} &\centering{$193493.0$} &\centering{$4000.0$} &&\centering{$193458$} &\centering{$193890.4$} &\centering{$4000.0$} &\\
\centering{RAND-150-2} &\centering{$\textbf{192597}$} &\centering{$\textbf{193032.6}$} &\centering{$4000.0$} &&\centering{$193066$} &\centering{$193386.2$} &\centering{$4000.0$} &&\centering{$193557$} &\centering{$194262.3$} &\centering{$4000.0$} &\\
\centering{RAND-150-3} &\centering{$\textbf{192439}$} &\centering{$\textbf{192913.4}$} &\centering{$4000.0$} &&\centering{$193151$} &\centering{$193406.8$} &\centering{$4000.0$} &&\centering{$193773$} &\centering{$194174.9$} &\centering{$4000.0$} &\\
\centering{RAND-200-1} &\centering{$\textbf{350294}$} &\centering{$\textbf{351403.0}$} &\centering{$12000.0$} &&\centering{$351535$} &\centering{$351853.7$} &\centering{$12000.0$} &&\centering{$352516$} &\centering{$352926.3$} &\centering{$12000.0$} &\\
\centering{RAND-200-2} &\centering{$\textbf{350995}$} &\centering{$\textbf{351602.4}$} &\centering{$12000.0$} &&\centering{$351698$} &\centering{$352355.9$} &\centering{$12000.0$} &&\centering{$352813$} &\centering{$353308.3$} &\centering{$12000.0$} &\\
\centering{RAND-200-3} &\centering{$\textbf{350929}$} &\centering{$\textbf{351404.1}$} &\centering{$12000.0$} &&\centering{$351488$} &\centering{$351896.4$} &\centering{$12000.0$} &&\centering{$351984$} &\centering{$352960.6$} &\centering{$12000.0$} &\\
\centering{RAND-250-1} &\centering{$\textbf{557649}$} &\centering{$\textbf{558353.2}$} &\centering{$20000.0$} &&\centering{$558516$} &\centering{$559169.6$} &\centering{$20000.0$} &&\centering{$560089$} &\centering{$560972.8$} &\centering{$20000.0$} &\\
\centering{RAND-250-2} &\centering{$\textbf{557044}$} &\centering{$\textbf{558775.7}$} &\centering{$20000.0$} &&\centering{$557795$} &\centering{$558927.6$} &\centering{$20000.0$} &&\centering{$559407$} &\centering{$560505.7$} &\centering{$20000.0$} &\\
\centering{RAND-250-3} &\centering{$\textbf{556806}$} &\centering{$\textbf{558307.0}$} &\centering{$20000.0$} &&\centering{$558184$} &\centering{$558753.5$} &\centering{$20000.0$} &&\centering{$559882$} &\centering{$560882.9$} &\centering{$20000.0$} &\\
\\
\centering{SOAK-150-1} &\centering{$\textbf{206721}$} &\centering{$\textbf{206898.8}$} &\centering{$4000.0$} &&\centering{$207024$} &\centering{$207305.0$} &\centering{$4000.0$} &&\centering{$206987$} &\centering{$207868.6$} &\centering{$4000.0$} &\\
\centering{SOAK-150-2} &\centering{$\textbf{206761}$} &\centering{$\textbf{207008.6}$} &\centering{$4000.0$} &&\centering{$207257$} &\centering{$207540.2$} &\centering{$4000.0$} &&\centering{$207616$} &\centering{$208000.0$} &\centering{$4000.0$} &\\
\centering{SOAK-150-3} &\centering{$\textbf{206759}$} &\centering{$\textbf{206880.0}$} &\centering{$4000.0$} &&\centering{$206926$} &\centering{$207223.2$} &\centering{$4000.0$} &&\centering{$206803$} &\centering{$207591.7$} &\centering{$4000.0$} &\\
\centering{SOAK-200-1} &\centering{$\textbf{369807}$} &\centering{$\textbf{370315.6}$} &\centering{$12000.0$} &&\centering{$370854$} &\centering{$371197.6$} &\centering{$12000.0$} &&\centering{$371647$} &\centering{$372112.1$} &\centering{$12000.0$} &\\
\centering{SOAK-200-2} &\centering{$\textbf{369878}$} &\centering{$\textbf{370254.3}$} &\centering{$12000.0$} &&\centering{$370507$} &\centering{$370960.9$} &\centering{$12000.0$} &&\centering{$370969$} &\centering{$371716.7$} &\centering{$12000.0$} &\\
\centering{SOAK-200-3} &\centering{$\textbf{369775}$} &\centering{$\textbf{370202.6}$} &\centering{$12000.0$} &&\centering{$370235$} &\centering{$371052.3$} &\centering{$12000.0$} &&\centering{$371011$} &\centering{$371984.3$} &\centering{$12000.0$} &\\
\centering{SOAK-250-1} &\centering{$\textbf{581536}$} &\centering{$\textbf{582299.2}$} &\centering{$20000.0$} &&\centering{$582927$} &\centering{$583364.0$} &\centering{$20000.0$} &&\centering{$583995$} &\centering{$584981.7$} &\centering{$20000.0$} &\\
\centering{SOAK-250-2} &\centering{$\textbf{581438}$} &\centering{$\textbf{582037.5}$} &\centering{$20000.0$} &&\centering{$582790$} &\centering{$583413.7$} &\centering{$20000.0$} &&\centering{$583976$} &\centering{$584647.1$} &\centering{$20000.0$} &\\
\centering{SOAK-250-3} &\centering{$\textbf{581819}$} &\centering{$\textbf{582384.6}$} &\centering{$20000.0$} &&\centering{$582787$} &\centering{$583752.2$} &\centering{$20000.0$} &&\centering{$584232$} &\centering{$585026.7$} &\centering{$20000.0$} &\\
\\
\centering{chr20a} &\centering{$2422$} &\centering{$2526.2$} &\centering{$899.0$} &&\centering{$\textbf{2192}$} &\centering{$2205.2$} &\centering{$543.7$} &&\centering{$\textbf{2192}$} &\centering{$\textbf{2192.8}$} &\centering{$503.1$} &\\
\centering{chr20b} &\centering{$2412$} &\centering{$2492.4$} &\centering{$707.9$} &&\centering{$\textbf{2298}$} &\centering{$\textbf{2363.2}$} &\centering{$560.5$} &&\centering{$\textbf{2298}$} &\centering{$2371.8$} &\centering{$507.7$} &\\
\centering{chr20c} &\centering{$16448$} &\centering{$17439.2$} &\centering{$982.6$} &&\centering{$\textbf{14142}$} &\centering{$\textbf{14142.0}$} &\centering{$395.0$} &&\centering{$\textbf{14142}$} &\centering{$\textbf{14142.0}$} &\centering{$298.4$} &\\
\centering{chr22a} &\centering{$6476$} &\centering{$6621.0$} &\centering{$1423.2$} &&\centering{$\textbf{6156}$} &\centering{$6158.0$} &\centering{$701.8$} &&\centering{$\textbf{6156}$} &\centering{$\textbf{6156.0}$} &\centering{$494.5$} &\\
\centering{chr22b} &\centering{$6538$} &\centering{$6664.8$} &\centering{$1167.1$} &&\centering{$\textbf{6194}$} &\centering{$6247.0$} &\centering{$952.8$} &&\centering{$\textbf{6194}$} &\centering{$\textbf{6237.2}$} &\centering{$624.0$} &\\
\centering{chr25a} &\centering{$4482$} &\centering{$4706.6$} &\centering{$1811.3$} &&\centering{$\textbf{3796}$} &\centering{$3894.0$} &\centering{$1483.8$} &&\centering{$\textbf{3796}$} &\centering{$\textbf{3803.8}$} &\centering{$857.8$} &\\
\centering{nug20} &\centering{$2688$} &\centering{$2715.4$} &\centering{$793.2$} &&\centering{$\textbf{2570}$} &\centering{$\textbf{2570.0}$} &\centering{$340.3$} &&\centering{$\textbf{2570}$} &\centering{$\textbf{2570.0}$} &\centering{$283.6$} &\\
\centering{nug21} &\centering{$2518$} &\centering{$2603.2$} &\centering{$1207.5$} &&\centering{$\textbf{2438}$} &\centering{$\textbf{2438.0}$} &\centering{$438.5$} &&\centering{$\textbf{2438}$} &\centering{$\textbf{2438.0}$} &\centering{$326.6$} &\\
\centering{nug22} &\centering{$3744$} &\centering{$3828.2$} &\centering{$1440.7$} &&\centering{$\textbf{3596}$} &\centering{$\textbf{3596.0}$} &\centering{$508.6$} &&\centering{$\textbf{3596}$} &\centering{$\textbf{3596.0}$} &\centering{$358.2$} &\\
\centering{nug24} &\centering{$3702$} &\centering{$3759.4$} &\centering{$2113.4$} &&\centering{$\textbf{3488}$} &\centering{$\textbf{3488.0}$} &\centering{$749.9$} &&\centering{$\textbf{3488}$} &\centering{$\textbf{3488.0}$} &\centering{$503.5$} &\\
\centering{nug25} &\centering{$4002$} &\centering{$4031.4$} &\centering{$2204.5$} &&\centering{$\textbf{3744}$} &\centering{$\textbf{3744.0}$} &\centering{$932.9$} &&\centering{$\textbf{3744}$} &\centering{$\textbf{3744.0}$} &\centering{$607.0$} &\\
\centering{nug27} &\centering{$5498$} &\centering{$5546.6$} &\centering{$2439.8$} &&\centering{$\textbf{5234}$} &\centering{$\textbf{5234.0}$} &\centering{$1209.9$} &&\centering{$\textbf{5234}$} &\centering{$\textbf{5234.0}$} &\centering{$827.1$} &\\
\centering{nug28} &\centering{$5422$} &\centering{$5479.2$} &\centering{$2550.3$} &&\centering{$\textbf{5166}$} &\centering{$\textbf{5166.0}$} &\centering{$1707.0$} &&\centering{$\textbf{5166}$} &\centering{$\textbf{5166.0}$} &\centering{$1083.7$} &\\
\centering{nug30} &\centering{$6506$} &\centering{$6621.0$} &\centering{$3776.0$} &&\centering{$\textbf{6124}$} &\centering{$6125.6$} &\centering{$2203.6$} &&\centering{$\textbf{6124}$} &\centering{$\textbf{6124.4}$} &\centering{$1783.6$} &\\

\hline
\end{tabular}
\end{scriptsize}
\end{table}
\renewcommand{\baselinestretch}{1.0}\small\normalsize

As mentioned in Section \ref{SubsecIntensiveSearch}, the proposed three-phase search algorithm relies on two tabu-based directed perturbation operators, using a parameter $p$ to control the probability for selecting each operator. To analyze the impact of different directed perturbation operators, we respectively let $p$ equal to 1 (using the swap-edge directed perturbation operator alone), 0.5 (combining two operators), 0 (using the swap-vertex directed perturbation operator alone), while keeping all the other ingredients and parameters unchanged as described in Section \ref{SecResults}, to get three variants of the proposed algorithm for comparisons.

For this experiment, we select a subset of 44 most challenging instances out of all the 659 instances, including the 12 instances with $n\ge100$ of group SS, all the 18 instances of groups RAND and SOAK, together with the 14 instances with $n\ge 40$ of group QAP-QMSTP, and then compare the performance of each variant based on these selected instances. For each instance, we independently run each variant 10 times, each run using the same terminal criterion like in Section \ref{SecResults}. The obtained results are provided in Table \ref{TblAnaylyzeDirectPerturb}, including the best and average cost of the 10 independent runs, as well as the accumulated CPU times (in seconds).

As shown in Table \ref{TblAnaylyzeDirectPerturb}, on one hand, for the general QMSTP instances of groups SS, RAND and SOAK, the variant with $p=1$ yields the best results on almost all (only with two exceptions) the test instances, in terms of both best and average costs, while consuming statistically much less time on group SS and the same cutoff time on groups RAND and SOAK. By contrast, the variant with $p=0$ performs the worst.

On the other hand, for the special instances transformed from QAP, we observe that the variants with $p=0.5$ and $p=0$ yield the same best results (with all instances solved to optimality), being much better than the ones obtained by the variant with $p=1$. Furthermore, considering the average cost and computation time, the variant with $p=0$ performs slightly better than the variant with $p=0.5$, and much better than the variant with $p=1$, indicating the importance of the swap-vertex based directed perturbation operator for solving these special instances.

Following these observations, in the standard TPS algorithm, we just use the swap-edge based directed perturbation operator alone ($p=1$) for solving the general QMSTP instances, and combine these two directed perturbation operators ($p=0.5$) together for solving the special instances transformed from QAP.

\subsection{Impact of the diversified perturbation operator}
\label{SubsecAnalyzeDiversifiedPerturb}

Now we turn our attention to the impact of the diversified perturbation operator. For this purpose, based on the standard TPS (denoted by TPS-V0 hereafter) algorithm described in Section \ref{SecApproach}, we implement as follows two compared variants by varying the diversified perturbation operator. Respectively, variant TPS-V1 adopts a random starting strategy which uses the randomized initialization procedure described in Section \ref{SubsecInit} instead of the original diversified perturbation operator in Section \ref{SubsecDiverSearch}, and variant TPS-V2 uses the directed perturbation operator in Section \ref{SubsecIntensiveSearch} instead. All the other ingredients and parameters keep in accordance with TPS-V0.

Again, we use the 44 most challenging instances of Section \ref{SubsecAnalyzeDirectedPerturb} as benchmarks to evaluate the performances of these compared TPS variants. For each instance, we also independently run each variant 10 times, each run using the same parameters and terminal criterion like in Section \ref{SecResults}. The results are detailed in Table \ref{TblAnaylyzeDiversifyPerturb}, with each column having the same meanings as in Table \ref{TblAnaylyzeDirectPerturb}.

\renewcommand{\baselinestretch}{0.75}\scriptsize\scriptsize
\begin{table}\centering
\begin{scriptsize}
\caption{Results corresponding to different diversified perturbation operators}
\label{TblAnaylyzeDiversifyPerturb}
\begin{tabular}{p{2cm}p{0.8cm}p{1.0cm}p{0.8cm}p{0cm}p{0.8cm}p{1.0cm}p{0.8cm}p{0cm}p{0.8cm}p{1.0cm}p{0.8cm}p{0cm}}
\hline
Instance &\multicolumn{3}{c}{TPS-V0} &&\multicolumn{3}{c}{TPS-V1} &&\multicolumn{3}{c}{TPS-V2} &\\
\cline{2-4} \cline{6-8} \cline{10-12}

&\centering{$Best$} &\centering{$Average$} &\centering{$t(s)$} &&\centering{$Best$} &\centering{$Average$} &\centering{$t(s)$} &&\centering{$Best$} &\centering{$Average$} &\centering{$t(s)$} &\\
\hline

\centering{SS-100-1} &\centering{$\textbf{88790}$} &\centering{$\textbf{89031.4}$} &\centering{$119.9$} &&\centering{$89031$} &\centering{$89189.6$} &\centering{$195.3$} &&\centering{$88939$} &\centering{$89138.6$} &\centering{$125.7$} &\\
\centering{SS-100-2} &\centering{$\textbf{88941}$} &\centering{$\textbf{89159.2}$} &\centering{$107.4$} &&\centering{$89071$} &\centering{$89200.2$} &\centering{$180.0$} &&\centering{$89008$} &\centering{$89229.5$} &\centering{$131.7$} &\\
\centering{SS-100-3} &\centering{$\textbf{88631}$} &\centering{$\textbf{88966.4}$} &\centering{$120.9$} &&\centering{$88815$} &\centering{$89082.4$} &\centering{$221.7$} &&\centering{$88944$} &\centering{$89116.9$} &\centering{$135.7$} &\\
\centering{SS-150-1} &\centering{$205201$} &\centering{$205757.0$} &\centering{$467.3$} &&\centering{$205553$} &\centering{$205897.2$} &\centering{$937.3$} &&\centering{$\textbf{205159}$} &\centering{$\textbf{205721.4}$} &\centering{$550.4$} &\\
\centering{SS-150-2} &\centering{$\textbf{205013}$} &\centering{$\textbf{205714.6}$} &\centering{$411.5$} &&\centering{$205498$} &\centering{$205927.8$} &\centering{$891.8$} &&\centering{$205476$} &\centering{$205844.6$} &\centering{$556.5$} &\\
\centering{SS-150-3} &\centering{$\textbf{204841}$} &\centering{$\textbf{205505.7}$} &\centering{$462.3$} &&\centering{$205514$} &\centering{$205921.8$} &\centering{$867.1$} &&\centering{$205411$} &\centering{$205720.1$} &\centering{$546.6$} &\\
\centering{SS-200-1} &\centering{$\textbf{371218}$} &\centering{$371827.7$} &\centering{$1105.9$} &&\centering{$371619$} &\centering{$372111.3$} &\centering{$2230.5$} &&\centering{$371306$} &\centering{$\textbf{371717.2}$} &\centering{$1423.7$} &\\
\centering{SS-200-2} &\centering{$371081$} &\centering{$\textbf{371556.2}$} &\centering{$1317.1$} &&\centering{$371421$} &\centering{$372103.2$} &\centering{$2587.1$} &&\centering{$\textbf{371011}$} &\centering{$371630.2$} &\centering{$1579.2$} &\\
\centering{SS-200-3} &\centering{$\textbf{371081}$} &\centering{$\textbf{371448.8}$} &\centering{$1524.6$} &&\centering{$371793$} &\centering{$372139.2$} &\centering{$2056.8$} &&\centering{$371320$} &\centering{$371775.6$} &\centering{$1419.0$} &\\
\centering{SS-250-1} &\centering{$\textbf{586445}$} &\centering{$\textbf{587322.1}$} &\centering{$2461.6$} &&\centering{$587344$} &\centering{$588184.3$} &\centering{$4689.6$} &&\centering{$586834$} &\centering{$587542.5$} &\centering{$5138.0$} &\\
\centering{SS-250-2} &\centering{$587210$} &\centering{$587526.1$} &\centering{$2193.8$} &&\centering{$587033$} &\centering{$588106.3$} &\centering{$5442.8$} &&\centering{$\textbf{586996}$} &\centering{$\textbf{587378.0}$} &\centering{$3663.7$} &\\
\centering{SS-250-3} &\centering{$586964$} &\centering{$587486.5$} &\centering{$2472.0$} &&\centering{$587442$} &\centering{$588066.0$} &\centering{$3764.7$} &&\centering{$\textbf{586728}$} &\centering{$\textbf{587203.3}$} &\centering{$4588.8$} &\\
\\
\centering{RAND-150-1} &\centering{$\textbf{192296}$} &\centering{$\textbf{192909.4}$} &\centering{$4000.0$} &&\centering{$193224$} &\centering{$193604.5$} &\centering{$4000.0$} &&\centering{$192444$} &\centering{$193027.5$} &\centering{$4000.0$} &\\
\centering{RAND-150-2} &\centering{$\textbf{192597}$} &\centering{$\textbf{193032.6}$} &\centering{$4000.0$} &&\centering{$193291$} &\centering{$193661.2$} &\centering{$4000.0$} &&\centering{$192834$} &\centering{$193159.5$} &\centering{$4000.0$} &\\
\centering{RAND-150-3} &\centering{$\textbf{192439}$} &\centering{$\textbf{192913.4}$} &\centering{$4000.0$} &&\centering{$193392$} &\centering{$193752.6$} &\centering{$4000.0$} &&\centering{$192725$} &\centering{$193019.8$} &\centering{$4000.0$} &\\
\centering{RAND-200-1} &\centering{$\textbf{350294}$} &\centering{$\textbf{351403.0}$} &\centering{$12000.0$} &&\centering{$351877$} &\centering{$352432.1$} &\centering{$12000.0$} &&\centering{$351035$} &\centering{$351517.5$} &\centering{$12000.0$} &\\
\centering{RAND-200-2} &\centering{$\textbf{350995}$} &\centering{$351602.4$} &\centering{$12000.0$} &&\centering{$351990$} &\centering{$352472.4$} &\centering{$12000.0$} &&\centering{$351215$} &\centering{$\textbf{351546.0}$} &\centering{$12000.0$} &\\
\centering{RAND-200-3} &\centering{$350929$} &\centering{$\textbf{351404.1}$} &\centering{$12000.0$} &&\centering{$352127$} &\centering{$352545.8$} &\centering{$12000.0$} &&\centering{$\textbf{350921}$} &\centering{$351447.2$} &\centering{$12000.0$} &\\
\centering{RAND-250-1} &\centering{$\textbf{557649}$} &\centering{$558353.2$} &\centering{$20000.0$} &&\centering{$559476$} &\centering{$559837.2$} &\centering{$20000.0$} &&\centering{$557701$} &\centering{$\textbf{558305.0}$} &\centering{$20000.0$} &\\
\centering{RAND-250-2} &\centering{$\textbf{557044}$} &\centering{$558775.7$} &\centering{$20000.0$} &&\centering{$558897$} &\centering{$559684.1$} &\centering{$20000.0$} &&\centering{$558016$} &\centering{$\textbf{558385.1}$} &\centering{$20000.0$} &\\
\centering{RAND-250-3} &\centering{$\textbf{556806}$} &\centering{$\textbf{558307.0}$} &\centering{$20000.0$} &&\centering{$559010$} &\centering{$559799.0$} &\centering{$20000.0$} &&\centering{$557946$} &\centering{$558583.9$} &\centering{$20000.0$} &\\
\\
\centering{SOAK-150-1} &\centering{$\textbf{206721}$} &\centering{$\textbf{206898.8}$} &\centering{$4000.0$} &&\centering{$207020$} &\centering{$207378.3$} &\centering{$4000.0$} &&\centering{$\textbf{206721}$} &\centering{$207070.8$} &\centering{$4000.0$} &\\
\centering{SOAK-150-2} &\centering{$\textbf{206761}$} &\centering{$\textbf{207008.6}$} &\centering{$4000.0$} &&\centering{$207204$} &\centering{$207579.2$} &\centering{$4000.0$} &&\centering{$207112$} &\centering{$207338.4$} &\centering{$4000.0$} &\\
\centering{SOAK-150-3} &\centering{$\textbf{206759}$} &\centering{$\textbf{206880.0}$} &\centering{$4000.0$} &&\centering{$207139$} &\centering{$207301.2$} &\centering{$4000.0$} &&\centering{$206777$} &\centering{$207054.4$} &\centering{$4000.0$} &\\
\centering{SOAK-200-1} &\centering{$\textbf{369807}$} &\centering{$\textbf{370315.6}$} &\centering{$12000.0$} &&\centering{$370849$} &\centering{$371570.2$} &\centering{$12000.0$} &&\centering{$370226$} &\centering{$370770.3$} &\centering{$12000.0$} &\\
\centering{SOAK-200-2} &\centering{$\textbf{369878}$} &\centering{$\textbf{370254.3}$} &\centering{$12000.0$} &&\centering{$370960$} &\centering{$371479.6$} &\centering{$12000.0$} &&\centering{$370056$} &\centering{$370599.2$} &\centering{$12000.0$} &\\
\centering{SOAK-200-3} &\centering{$\textbf{369775}$} &\centering{$\textbf{370202.6}$} &\centering{$12000.0$} &&\centering{$370943$} &\centering{$371233.8$} &\centering{$12000.0$} &&\centering{$370279$} &\centering{$370694.4$} &\centering{$12000.0$} &\\
\centering{SOAK-250-1} &\centering{$\textbf{581536}$} &\centering{$\textbf{582299.2}$} &\centering{$20000.0$} &&\centering{$583223$} &\centering{$584110.7$} &\centering{$20000.0$} &&\centering{$581955$} &\centering{$582939.1$} &\centering{$20000.0$} &\\
\centering{SOAK-250-2} &\centering{$\textbf{581438}$} &\centering{$\textbf{582037.5}$} &\centering{$20000.0$} &&\centering{$583474$} &\centering{$583982.8$} &\centering{$20000.0$} &&\centering{$582285$} &\centering{$582827.2$} &\centering{$20000.0$} &\\
\centering{SOAK-250-3} &\centering{$\textbf{581819}$} &\centering{$\textbf{582384.6}$} &\centering{$20000.0$} &&\centering{$583178$} &\centering{$584170.3$} &\centering{$20000.0$} &&\centering{$582838$} &\centering{$583190.3$} &\centering{$20000.0$} &\\

\\
\centering{chr20a} &\centering{$\textbf{2192}$} &\centering{$\textbf{2205.2}$} &\centering{$543.7$} &&\centering{$\textbf{2192}$} &\centering{$2208.0$} &\centering{$549.4$} &&\centering{$\textbf{2192}$} &\centering{$2208.8$} &\centering{$549.4$} &\\
\centering{chr20b} &\centering{$\textbf{2298}$} &\centering{$\textbf{2363.2}$} &\centering{$560.5$} &&\centering{$2352$} &\centering{$2366.2$} &\centering{$579.6$} &&\centering{$\textbf{2298}$} &\centering{$\textbf{2363.2}$} &\centering{$539.4$} &\\
\centering{chr20c} &\centering{$\textbf{14142}$} &\centering{$\textbf{14142.0}$} &\centering{$395.0$} &&\centering{$\textbf{14142}$} &\centering{$\textbf{14142.0}$} &\centering{$388.5$} &&\centering{$\textbf{14142}$} &\centering{$\textbf{14142.0}$} &\centering{$409.1$} &\\
\centering{chr22a} &\centering{$\textbf{6156}$} &\centering{$6158.0$} &\centering{$701.8$} &&\centering{$\textbf{6156}$} &\centering{$6158.0$} &\centering{$626.9$} &&\centering{$\textbf{6156}$} &\centering{$\textbf{6156.0}$} &\centering{$781.9$} &\\
\centering{chr22b} &\centering{$\textbf{6194}$} &\centering{$\textbf{6247.0}$} &\centering{$952.8$} &&\centering{$6230$} &\centering{$6257.2$} &\centering{$830.1$} &&\centering{$6230$} &\centering{$6248.2$} &\centering{$941.4$} &\\
\centering{chr25a} &\centering{$\textbf{3796}$} &\centering{$3894.0$} &\centering{$1483.8$} &&\centering{$\textbf{3796}$} &\centering{$3892.0$} &\centering{$1416.2$} &&\centering{$\textbf{3796}$} &\centering{$\textbf{3841.2}$} &\centering{$1433.6$} &\\
\centering{nug20} &\centering{$\textbf{2570}$} &\centering{$\textbf{2570.0}$} &\centering{$340.3$} &&\centering{$\textbf{2570}$} &\centering{$\textbf{2570.0}$} &\centering{$343.7$} &&\centering{$\textbf{2570}$} &\centering{$\textbf{2570.0}$} &\centering{$368.2$} &\\
\centering{nug21} &\centering{$\textbf{2438}$} &\centering{$\textbf{2438.0}$} &\centering{$438.5$} &&\centering{$\textbf{2438}$} &\centering{$\textbf{2438.0}$} &\centering{$434.6$} &&\centering{$\textbf{2438}$} &\centering{$\textbf{2438.0}$} &\centering{$447.0$} &\\
\centering{nug22} &\centering{$\textbf{3596}$} &\centering{$\textbf{3596.0}$} &\centering{$508.6$} &&\centering{$\textbf{3596}$} &\centering{$\textbf{3596.0}$} &\centering{$487.4$} &&\centering{$\textbf{3596}$} &\centering{$\textbf{3596.0}$} &\centering{$501.3$} &\\
\centering{nug24} &\centering{$\textbf{3488}$} &\centering{$\textbf{3488.0}$} &\centering{$749.9$} &&\centering{$\textbf{3488}$} &\centering{$\textbf{3488.0}$} &\centering{$638.8$} &&\centering{$\textbf{3488}$} &\centering{$\textbf{3488.0}$} &\centering{$658.4$} &\\
\centering{nug25} &\centering{$\textbf{3744}$} &\centering{$\textbf{3744.0}$} &\centering{$932.9$} &&\centering{$\textbf{3744}$} &\centering{$\textbf{3744.0}$} &\centering{$886.5$} &&\centering{$\textbf{3744}$} &\centering{$\textbf{3744.0}$} &\centering{$787.2$} &\\
\centering{nug27} &\centering{$\textbf{5234}$} &\centering{$\textbf{5234.0}$} &\centering{$1209.9$} &&\centering{$\textbf{5234}$} &\centering{$\textbf{5234.0}$} &\centering{$1234.8$} &&\centering{$\textbf{5234}$} &\centering{$\textbf{5234.0}$} &\centering{$1228.4$} &\\
\centering{nug28} &\centering{$\textbf{5166}$} &\centering{$\textbf{5166.0}$} &\centering{$1707.0$} &&\centering{$\textbf{5166}$} &\centering{$\textbf{5166.0}$} &\centering{$1705.0$} &&\centering{$\textbf{5166}$} &\centering{$\textbf{5166.0}$} &\centering{$1610.7$} &\\
\centering{nug30} &\centering{$\textbf{6124}$} &\centering{$6125.6$} &\centering{$2203.6$} &&\centering{$\textbf{6124}$} &\centering{$\textbf{6124.4}$} &\centering{$2736.6$} &&\centering{$\textbf{6124}$} &\centering{$\textbf{6124.4}$} &\centering{$2434.5$} &\\

\hline
\end{tabular}
\end{scriptsize}
\end{table}
\renewcommand{\baselinestretch}{1.0}\small\normalsize

As shown in Table \ref{TblAnaylyzeDiversifyPerturb}, we observe that TPS-V0 yields better results than TPS-V2 on 8 out of the 12 SS instances, and dominates TPS-V1 on all the 12 SS instances, while consuming statistically much less computation time with respect to these two compared variants. Furthermore, for the  groups RAND and SOAK, TPS-V0 performs clearly much better than the two compared variants in terms of both best and average solution quality, while consuming the same computation time. Finally, for the special instances transformed from the QAP instances, TPS-V0 yields slightly better results than the two other variants, without a significant statistical difference, in terms of both solution quality and runtime. This experiment clearly confirms the importance of the diversified perturbation operator to the proposed algorithm, especially for solving conventional QMSTP instances.

\section{Conclusion}
\label{SecConclude}

We have originally proposed a three-phase heuristic algorithm named TPS for the \textsl{quadratic minimum spanning tree problem (QMSTP)}, which could be used to model a number of network designing problems. TPS consists of a descent-based neighborhood search phase for local optimization, a local optima exploring phase for intensive search in a given regional search space, a diversified perturbation phase for jumping out of the current regional search space. TPS integrates a fast evaluation technique to avoid useless computations. An extensive experimental comparison on all the available benchmarks shows that TPS produces highly competitive results with respect to the state-of-the-art heuristics. For the 630 conventional QMSTP instances, TPS succeeds in discovering improved best known solutions for 31 challenging instances, while for almost all (only with two exceptions) the remaining instances, TPS matches the best known results, with a reasonable computational time. In particular, for all the 29 special instances transformed from the QAP problem, TPS can unexceptionally reach the optimal solutions within a short time, while the previous best QMSTP algorithm can only solve 12 out of these 29 instances to optimality with much more computing times.

Furthermore, we analyze the influences of several important ingredients of the proposed algorithm. Experimental results based on a subset of selected most challenging instances show that: (1) The fast examination technique used in the first search phase is able to identify and discard a large number of useless moves, and thus contributes significantly to the efficiency of the proposed algorithm. (2) The directed perturbation operators used in the second search phase are extremely important to the search process, specifically, the swap-edge based directed perturbation operator is suitable for the  conventional QMSTP instances, while the swap-vertex based directed perturbation operator is particularly important for the special instances transformed from the QAP instances. (3) The diversified perturbation operator plays an important role to the proposed algorithm, especially for solving the conventional QMSTP instances.

\section*{Acknowledgements}
The work is partially supported by the Pays de la Loire Region within the RaDaPop (2009-2013) and LigeRO (2010-2014) projects, and the PGMO project (2013-2015). Sincere thanks to Dr. Roberto Cordone for kindly making their instances available and answering our questions.

\bibliographystyle{plain}

\end{document}